\documentclass{aa}
\usepackage[varg]{txfonts}
\usepackage{natbib}
\usepackage{graphicx}
\usepackage{mathabx}
\usepackage{amsmath}
\usepackage[normalem]{ulem}

\usepackage{color}
\usepackage{multirow}
\usepackage{longtable}
\usepackage{pdflscape}

\bibpunct{(}{)}{;}{a}{}{,}

\begin{document}

\title{The \textit{\textbf{Gaia}}-ESO Survey and CSI~2264: Substructures, disks, and sequential star formation in the young open cluster NGC~2264\thanks{Table~\ref{tab:data} is only available in electronic form at the CDS via anonymous ftp to cdsarc.u-strasbg.fr (130.79.128.5) or via http://cdsweb.u-strasbg.fr/cgi-bin/qcat?J/A+A/}}

\author{L. Venuti\inst{1}, L. Prisinzano\inst{1}, G.~G. Sacco\inst{2}, E. Flaccomio\inst{1}, R. Bonito\inst{1}, F. Damiani\inst{1}, G. Micela\inst{1}, M.~G. Guarcello\inst{1}, S. Randich\inst{2}, J.~R. Stauffer\inst{3}, A.~M. Cody\inst{4}, R.~D. Jeffries\inst{5}, S.~H.~P. Alencar\inst{6}, E.~J. Alfaro\inst{7}, A.~C. Lanzafame\inst{8,9}, E. Pancino\inst{2}, A. Bayo\inst{10}, G. Carraro\inst{11}, M.~T. Costado\inst{7}, A. Frasca\inst{9}, P. Jofr\'e\inst{12,13}, L. Morbidelli\inst{2}, S.~G. Sousa\inst{14}, S. Zaggia\inst{15}}

\institute{INAF -- Osservatorio Astronomico di Palermo G.\,S. Vaiana, Piazza del Parlamento 1, 90134 Palermo, Italy\\ e-mail: lvenuti@astropa.unipa.it \and INAF -- Osservatorio Astrofisico di Arcetri, Largo E. Fermi 5, 50125, Florence, Italy \and Spitzer Science Center (SSC), California Institute of Technology, Pasadena, CA 91125, USA \and NASA Ames Research Center, Moffett Field, CA 94035, USA \and Astrophysics Group, Keele University, Keele, Staffordshire ST5 5BG, United Kingdom \and Departamento de F\'isica – ICEx – UFMG, Av. Ant\^onio Carlos 6627, 30270-901 Belo Horizonte, MG, Brazil \and Instituto de Astrof\'{i}sica de Andaluc\'{i}a-CSIC, Apdo. 3004, 18080, Granada, Spain \and Universit\`a di Catania, Dipartimento di Fisica e Astronomia, Sezione Astrofisica, Via S. Sofia 78, 95123 Catania, Italy \and INAF - Osservatorio Astrofisico di Catania, via S. Sofia 78, 95123, Catania, Italy \and Instituto de F\'isica y Astronom\'ia, Universidad de Valpara\'iso, Chile \and Dipartimento di Fisica e Astronomia, Universit\`a di Padova, Vicolo dell'Osservatorio 3, 35122 Padova, Italy \and Institute of Astronomy, University of Cambridge, Madingley Road, Cambridge CB3 0HA, United Kingdom \and N\'ucleo de Astronom\'{i}a, Facultad de Ingenier\'{i}a, Universidad Diego Portales, Av. Ej\'ercito 441, Santiago, Chile \and Instituto de Astrof\'isica e Ci\^encias do Espa\c{c}o, Universidade do Porto, CAUP, Rua das Estrelas, 4150-762 Porto, Portugal \and INAF - Padova Observatory, Vicolo dell'Osservatorio 5, 35122 Padova, Italy}

\date{Received 4 May 2017 / Accepted 5 September 2017}

\abstract {Reconstructing the structure and history of young clusters is pivotal to understanding the mechanisms and timescales of early stellar evolution and planet formation. Recent studies suggest that star clusters often exhibit a hierarchical structure, possibly resulting from several star formation episodes occurring sequentially rather than a monolithic cloud collapse.}{We aim to explore the structure of the open cluster and star-forming region NGC~2264 ($\sim$3~Myr), which is one of the youngest, richest and most accessible star clusters in the local spiral arm of our Galaxy; we link the spatial distribution of cluster members to other stellar properties such as age and evolutionary stage to probe the star formation history within the region.}{We combined spectroscopic data obtained as part of the {\it Gaia}-ESO Survey (GES) with multi-wavelength photometric data from the Coordinated Synoptic Investigation of NGC~2264 (CSI~2264) campaign. We examined a sample of 655 cluster members, with masses between 0.2 and 1.8~M$_\odot$ and including both disk-bearing and disk-free young stars. We used $T_\mathrm{eff}$ estimates from GES and $g,r,i$ photometry from CSI~2264 to derive individual extinction and stellar parameters.}{We find a significant age spread of 4--5~Myr among cluster members. Disk-bearing objects are statistically associated with younger isochronal ages than disk-free sources. The cluster has a hierarchical structure, with two main blocks along its latitudinal extension. The northern half develops around the O-type binary star S~Mon; the southern half, close to the tip of the Cone Nebula, contains the most embedded regions of NGC~2264, populated mainly by objects with disks and ongoing accretion. The median ages of objects at different locations within the cluster, and the spatial distribution of disked and non-disked sources, suggest that star formation began in the north of the cluster, over 5~Myr ago, and was ignited in its southern region a few Myr later. Star formation is likely still ongoing in the most embedded regions of the cluster, while the outer regions host a widespread population of more evolved objects; these may be the result of an earlier star formation episode followed by outward migration on timescales of a few Myr. We find a detectable lag between the typical age of disk-bearing objects and that of accreting objects in the inner regions of NGC~2264: the first tend to be older than the second, but younger than disk-free sources at similar locations within the cluster. This supports earlier findings that the characteristic timescales of disk accretion are shorter than those of disk dispersal, and smaller than the average age of NGC~2264 (i.e., $\lesssim$3~Myr). At the same time, we note that disks in the north of the cluster tend to be shorter-lived ($\sim$2.5~Myr) than elsewhere; this may reflect the impact of massive stars within the region (notably S~Mon), that trigger rapid disk dispersal.}{Our results, consistent with earlier studies on NGC~2264 and other young clusters, support the idea of a star formation process that takes place sequentially over a prolonged span in a given region. A complete understanding of the dynamics of formation and evolution of star clusters requires accurate astrometric and kinematic characterization of its population; significant advance in this field is foreseen in the upcoming years thanks to the ongoing \emph{Gaia} mission, coupled with extensive ground-based surveys like GES.}


\titlerunning{GES and CSI~2264: Substructures, disks, and sequential star formation in NGC~2264}
\authorrunning{L.~Venuti et al.}

\maketitle

\section{Introduction} \label{sec:introduction}

Characterizing the properties and dynamical evolution of stellar clusters is a crucial aspect for studies of star formation. It is estimated that a significant fraction of stars are formed in embedded clusters; these can count tens to hundreds of members, spanning a variety of evolutionary stages \citep[see][]{lada2003, kennicutt2012}. Young clusters are important laboratories in which to test scenarios of star and planet formation. As an example, measuring the fraction of stars in different evolutionary stages as a function of the average cluster age provides unique observational constraints on the timescales inherent to the early stellar evolution and protoplanetary disk survival, vital to inform credible theoretical models of star formation and planet building in the circumstellar disks.

Significant advance in our empirical knowledge of the structure of star clusters has been marked, over the last two decades, by the advent of large-scale surveys and imaging cameras in the infrared (most notably the Two Micron All Sky Survey, 2MASS, \citealp{2MASS}, and the Infrared Array Camera onboard the {\it Spitzer} Space Telescope, IRAC, \citealp{fazio2004}). These enabled the first systematic surveys to identify and map embedded clusters and their stellar content, not accessible at optical wavelengths due to the strong extinction by the intervening material \citep{lada2010}. Such studies have revealed that young clusters may exhibit broadly different morphologies \citep{allen2007}: while some appear compact and spherical, several display an elongated and filamentary shape, reminiscent of that of molecular clouds in which they are born. A hierarchical or fractal structure, with multiple subclusterings, is another common feature among young embedded clusters \citep[e.g.,][]{schmeja2006, elmegreen2010, kuhn2014}. This may reflect a formation process that occurs in a bottom-up fashion, where many small subclusters form along filamentary structures of the molecular cloud, grow by accreting other stars, and eventually merge together to shape the final, large-scale cluster \citep{bonnell2003, clarke2010}. 

Knowing the median age and the age dispersion in the population of a young cluster is of great interest to understand the duration and modes of star formation within the region. The age spread, in particular, bears information on whether all cluster members were born in a single, short-lived star formation burst \citep[e.g.,][]{kudryavtseva2012}, or instead resulted from a continuous or sequential star formation process that stretched over a longer time span \citep[e.g.,][]{preibisch2007}. Prolonged star formation activity may reflect a scenario in which local star formation in dense cores is driven by large-scale, quasi-static contraction of the parent cloud in the presence of a magnetic field \citep{palla2002, palla2004}; in contrast, evidence of rapid star formation may favor a scenario regulated by the interplay between turbulence and gravity on dynamical timescales \citep[e.g.,][]{elmegreen2000, maclow2004}. Individual stellar ages are most straightforwardly estimated by plotting the objects on a Hertzsprung-Russell (HR) diagram\footnote{\citet{stahler_palla} argue that this approach is problematic for embedded protostars, whose observed spectra are significantly impacted by the surrounding dusty envelopes and may not be entirely representative of the physical properties of the central sources. A number of studies (see \citealp{white2007} and references therein) have reported measurements of protostars' fundamental properties ($T_{\rm eff}$, L$_{bol}$) from scattered light through the outflow cavities in their circumstellar envelopes; however, the high extinctions typical of these objects limit the statistics for these sources. On the other hand, a (conceptually) straightforward placing of large samples of young stars on the HR diagram can be achieved starting from the pre-main sequence phase of early stellar evolution, when the sources have become optically visible.}, and then comparing the empirical cluster locus on the diagram to theoretical isochrones extracted from pre-main sequence (PMS) evolutionary models \citep[e.g.,][]{siess2000, baraffe2015}. Several studies have applied this method to various embedded clusters to report age spreads of the order of the average age of the region (a few Myr; see, e.g., \citealp{lada2003} and references therein). This would suggest that molecular clouds can typically sustain a star formation activity over several Myr. However, these results rely heavily on the accuracy of the locations of individual objects on a HR diagram, which can be severely affected by many unrelated sources of observational uncertainty, such as unresolved binaries, disk orientation, stellar variability, extinction, or the specific accretion history of individual objects \citep[e.g.,][]{hartmann2001, baraffe2009, soderblom2010, preibisch2012}. As a consequence, the age dispersion derived from isochrone-fitting would only provide a (model-dependent) upper limit to the true age spread within the cluster, and additional information on the nature of the objects is required to confirm or disprove such spread (see, e.g., \citealp{palla2005}).

Mapping the kinematic properties of young stars across a given cluster is an efficient tracer of the structure, dynamics, and star formation history of the region. First systematic surveys of the radial velocity across young populous clusters such as the Orion Nebula Cluster \citep[ONC;][]{furesz2008, tobin2009, dario2017} or NGC~2264 \citep{furesz06, tobin2015} have revealed a structured spatial distribution of the kinematic properties of cluster members, that follows that of the molecular gas in the region. Similar studies conducted on slightly older clusters (e.g., \citealp{jeffries2006} on $\sigma$~Ori, $\sim$6~Myr; \citealp{bell2013}) have also reported the presence of two or more kinematically distinct components among cluster members, and suggested that these may correspond to multiple populations of different age within the region. Extensive and homogeneous investigations of the kinematics of young stellar populations in different environments and evolutionary stages are needed in order to assemble these intra-cluster studies into a comprehensive picture of star cluster formation and evolution. This is a major motivation for the star cluster component of the currently ongoing {\it Gaia}-ESO Survey \citep[GES;][]{gilmore2012, randich2013}.

Launched at the end of 2011 and with an observing schedule running over six years, the GES spectroscopic survey has targeted about sixty open star clusters, among which about ten younger than $\sim$10~Myr. Data products from the campaign include astrophysical parameters and metallicities, element abundances and radial velocities for thousands of stars in our Galaxy. First published results on the kinematic properties of GES clusters (\citealp{jeffries2014} in $\gamma^2$~Vel; \citealp{sacco2015} in NGC~2547; \citealp{rigliaco2016} in L1688; \citealp{sacco2017} in Cha~I) support the view that the presence of multiple populations is indeed a common feature of young clusters and star-forming regions (SFRs). This, in turn, supports the view that clusters are not originated in a monolithic collapse process but in the merger of several substructures that have formed and evolved separately.

Among the clusters observed during the GES campaign, NGC~2264 ($\sim$3~Myr; see \citealp{dahm08} for a review) is one of the youngest. The few hundred publications existing to date on this SFR, and distributed over the last seventy years, attest to the relevance of NGC~2264 to the scenario of star formation. NGC~2264 is in fact one of the richest and most accessible young clusters in our Galaxy. It encompasses over 1\,000 confirmed members, that span all PMS evolutionary stages \citep{lada1987}: from protostellar sources still embedded in the parent cloud (Class~0/I), to PMS stars surrounded by an optically thick disk (Class~II), to finally disk-free young stars (Class~III). Moreover, in spite of being located at a distance of 760~pc \citep{sung97}, NGC~2264 suffers very little foreground extinction, and the associated molecular cloud complex significantly reduces the contamination from background stars. Recent studies (\citealp{sung08, sung09} and references therein) have shown that the distribution of NGC~2264 members is not uniform across the region: the cluster exhibits two separate, embedded cores with evidence of ongoing star formation activity, surrounded by a halo of scattered, more evolved members. 

Among the various observing campaigns that have targeted the region at all wavelengths over the past decades, the Coordinated Synoptic Investigation of NGC~2264 \citep[CSI~2264;][]{cody2014} is one of the most ambitious. This campaign, launched in December 2011 for over two months, provided a first synoptic characterization and simultaneous monitoring of hundreds of young stars at infrared, optical, UV and X-ray wavelengths on timescales varying from a few hours to several weeks. The extensive photometric and spectroscopic information gathered for NGC~2264 members during the CSI~2264 and GES campaigns constitutes a unique dataset available to date for a single SFR, in terms of completeness and accuracy. In this paper, we combine results from the GES and CSI~2264 surveys to investigate the nature of the PMS population of NGC~2264 and probe its dynamics and star formation history. 

The paper is organized as follows. In Sect.~\ref{sec:observations}, we introduce GES, CSI~2264, and the data relevant for this work. In Sect.\,\ref{sec:sample_selection}, we discuss the criteria used to identify cluster members among GES targets in the NGC~2264 field; we evaluate the completeness of our sample of members and assess what fraction of these objects exhibits signatures of circumstellar disk and ongoing accretion. In Sect.\,\ref{sec:results}, we derive extinction, bolometric luminosity, mass and age for objects in our sample; we note the apparent age spread on the HR diagram of the cluster, and compare this information with disk and accretion properties and with different age tracers to assess the veracity of such a spread. In Sect.\,\ref{sec:structure_SFH}, we explore the spatial distribution of Class~II/Class~III and accreting vs. non-accreting objects separately and as a function of age, to probe the structure and star formation history of the region. Our conclusions are summarized in Sect.\,\ref{sec:conclusions}.

\section{Observations} \label{sec:observations}

This work is based on observations acquired and reduced as part of two international observing campaigns: GES and CSI~2264. These two projects are briefly addressed and the relevant data used in this paper presented in the following Sections.

\subsection{Data from GES}

GES has been devised as an extensive spectroscopic survey of about 100\,000 objects, which primarily consist of Galactic field and open cluster stars. Performed with the GIRAFFE and UVES instruments of the FLAMES spectrograph at the Very Large Telescope (VLT), the survey aims at providing a homogeneous overview of the kinematics and chemical abundances distribution across the Milky Way. Data products from this program will complement the accurate astrometric survey of the Galaxy that is the core of the \emph{Gaia} mission \citep{gaia2016} launched by the European Space Agency (ESA), with the final goal of building a three-dimensional map of the Milky Way and delineating its structure, formation history and dynamical evolution. 

The GES observations of NGC~2264 were performed in three observing runs between October 2012 and January 2013. Spectroscopic parameters adopted in this paper are those provided in the 4th internal data release from the GES campaign (GESiDR4\footnote{Data extracted from the GES science archive at http://ges.roe.ac.uk/. GESiDR4 is also the base for the most recent GES public data release, available at the ESO archive (http://archive.eso.org).}), the most recent available at the time of the analysis. GESiDR4 contains entries for 1892 objects in the NGC~2264 field, over 80\% of which were observed using GIRAFFE. Targets were selected following the common guidelines adopted for young open clusters within the GES consortium (Bragaglia et al., in preparation), to ensure homogeneity across the dataset acquired for the project. In short, a preliminary list of confirmed cluster members was built upon several literature studies \citep[e.g.,][]{rebull2002, dahm05, flaccomio06, sung08, sung09} and criteria including IR excess, X-ray emission, H$\alpha$ emission, spatial distribution. This ensemble of objects was used to trace the cluster locus on various color-magnitude diagrams (CMDs; V vs. V-I and J vs. J-H); then, the final list of potential GIRAFFE targets was assembled by extracting, from the CMDs, all sources that fell onto the identified cluster locus. This strategy, based on photometric membership, was designed with the goal of obtaining a sample of candidate members as unbiased and as complete as possible. Conversely, UVES targets were selected very carefully among high-confidence cluster members, to meet the scientific goals of deriving accurate metallicity and chemical abundances.

The final input catalog of NGC~2264 comprised 2650 candidate members. This catalog was then used to schedule the observations (telescope pointing and fiber allocation), in such a way as to ensure an even coverage of the region across its whole spatial extent and, at the same time, allow an efficient use of the spectrograph fibers. Data were finally acquired for 1892 of the 2650 sources in the starting input catalog. A small fraction of targets ($\sim$7\%) already possessed archival FLAMES spectra (most notably from the CSI~2264 campaign, presented in Sect.\,\ref{sec:CSI2264}), and were not re-observed during the GES run on NGC~2264; raw spectra for these objects were retrieved from the ESO archive and included in the GES database, to be reduced homogeneously with spectra obtained for GES targets. The reduction of raw spectra was handled at the Cambridge Astronomy Survey Unit (CASU) for data acquired with GIRAFFE \citep{jeffries2014}, and in Arcetri for data acquired with UVES \citep{sacco2014}. Astrophysical parameters and chemical abundances were then extracted from the reduced spectra independently by several working groups (WGs) within the GES consortium (\citealp{smiljanic2014}; \citealp{lanzafame2015}; Recio-Blanco et al., in prep.), and were submitted to the Survey Parameter Homogenisation WG before being released\footnote{See https://www.gaia-eso.eu/ for details on the survey management within the consortium.} (Hourihane et al., in prep.).

In this work, we make use of the following GESiDR4 parameters for NGC~2264 targets:

\begin{itemize}
\item {\it EW(Li)}, equivalent width of the Li I 6708\,\AA\mbox{ }line, as indicator of youth \citep[e.g.,][]{soderblom2010, soderblom2014};
\item {\it EW(H$\alpha$)} and {\it W10\%(H$\alpha$)}, the equivalent width and width at 10\% intensity of the H$\alpha$ line, as proxies of mass accretion onto the central object \citep[e.g.,][]{white_basri2003};
\item {\it $\gamma$-index} \citep{damiani2014}, a spectral index sensitive to stellar gravity;
\item {\it $T_{\rm eff}$}, effective temperature of the stars \citep{lanzafame2015}.
\end{itemize}

\subsection{Data from CSI~2264} \label{sec:CSI2264}

CSI~2264 \citep{cody2014} consisted of an unprecedented cooperative effort aimed at characterizing the multi-wavelength variability of young stars in NGC~2264 simultaneously from the X-rays to the mid-IR, on timescales ranging from fractions of hours to several weeks. The bulk of the observing campaign was carried out between December 2011 and February 2012. The {\it Spitzer} and {\it CoRoT} space telescopes provided the backbone of the infrared and optical photometric observations, respectively; they were employed to monitor the region simultaneously over a window of 30 consecutive days, providing light curves with photometric accuracy of $\lesssim$\,1\% for several hundred cluster members. Thirteen additional space-based and ground-based telescopes were employed in synergy with {\it Spitzer} and {\it CoRoT} to complement them with other spectral windows. In particular, the wide-field imaging facility MegaPrime/MegaCam at the Canada-France-Hawaii Telescope (CFHT) was used to map the cluster in the optical and in the UV ($u$, $g$, $r$ and $i$ filters of the Sloan Digital Sky Survey photometric system, SDSS, \citealp{F96SDSS}), and to monitor the variability of cluster members simultaneously in the optical and in the UV to derive direct diagnostics of accretion onto the stars (i.e., the UV excess; see \citealp{venuti2014} for details). The unique nature of the dataset gathered for the campaign renders NGC~2264, the only young open cluster for which such multi-wavelength and multi-instrument observations could be conceived, a landmark in investigations of the dynamic environment of young stars and their disks \citep{C13CSI2264}. 

The CSI~2264 photometric database was used here to complement the spectroscopic information available for GES targets in the NGC~2264 field. Sources in the GES catalog for NGC~2264 were identified with their CSI~2264 counterparts based on their spatial coordinates. The cross-correlation between the two catalogs was performed with TOPCAT \citep{TOPCAT} using a matching radius of 1~arcsec and retaining only the best (i.e., closest) match for each entry. The following information from the CSI~2264 database was used for the present analysis:

\begin{itemize}
\item $g,\,r,\,i$ photometry from the CFHT observations \citep{venuti2014};
\item UV excess information from the CFHT observations \citep{venuti2014}, to classify accreting and non-accreting sources;
\item IR excess information from {\it Spitzer}/IRAC observations \citep{cody2014}, to classify disk-bearing and disk-free sources.
\end{itemize}

\noindent Additional photometric data from previous surveys conducted on the NGC~2264 region was also used in some parts of the study.

\section{Sample selection} \label{sec:sample_selection}

\subsection{NGC~2264 members in the GES catalog} \label{sec:memb_selection}

To identify cluster members among GES targets, we proceeded as follows.

A first selection was performed based on spectroscopic criteria and GES parameters (Sacco et al., in preparation), following the same procedure presented in \citet{sacco2017} for Cha~I. In a first step, the $\gamma$-index was used to separate PMS stars and dwarfs from giants in the sample and reject the latter: namely, as illustrated in \citet{damiani2014} (see also \citealp{prisinzano2016}), giants were identified as objects with $T_{\rm eff} < 5\,600$~K and $\gamma > 1.0$. Then, the EW(Li) and W10\%(H$\alpha$) parameters, respectively indicators of youth \citep{soderblom2010} and of disk accretion \citep{white_basri2003}, were used to select the sources in our sample that qualify as PMS stars. We decided to not include the radial velocities (RV) in our spectroscopic analysis of membership because previous studies \citep{furesz06, tobin2015} have shown that the RV distribution of NGC~2264 is non-Gaussian but exhibits several substructures and a significant dispersion.

The procedure described above allowed us to classify as cluster members 567 of the 1892 GES targets in the NGC~2264 field. As for the remaining 1325 stars, while in several cases the spectroscopic information available disqualifies them as members, for others ($\sim$500) no sufficient information is available from GES to either confirm or discard their belonging to the cluster (due to, e.g., low S/N in the spectra that prevents accurate measurements of EW(Li)). We therefore complemented the membership analysis of the GES sample by including also photometry and other information available from the CSI~2264 catalog\footnote{Open access at https://irsa.ipac.caltech.edu/data/SPITZER/CSI2264/} and database. This gathers all detected sources in the NGC~2264 region from either the CSI~2264 campaign or a number of previous investigations in different wavelength domains \citep{rebull2002, lamm04, ramirez04, dahm05, flaccomio06, furesz06, sung08, sung09}. A membership flag is assigned to each source listed in the CSI~2264 catalog based on several criteria, including photometric or spectroscopic H$\alpha$ emission, IR excess, photometric or kinematic properties consistent with the cluster locus, X-ray emission (see \citealp{cody2014} for further details). Objects that exhibit strong membership signatures on two or more of these criteria are assigned a membership flag of 1 (``very likely member''); those that satisfy only one of these criteria are flagged as 2 (``possible member''); membership flags of -1 or 0 indicate respectively ``likely field objects'' or objects with ``no membership information''. 

To ensure that the GES and the CSI~2264 member classification schemes are consistent with each other, we first examined the membership flags assigned in the CSI~2264 catalog to sources that were selected as cluster members based on GES spectroscopic parameters. All but three of the 567 GES targets classified as NGC~2264 members have a counterpart in the CSI~2264 catalog. 82\% of them resulted to also be classified as very likely members according to the criteria adopted within the CSI~2264 survey, and another 17\% were listed as possible members; only for three objects did the classification as members derived from GES data conflict with the classification as likely field objects in the CSI~2264 catalog. However, a close inspection of these three objects in the CSI~2264 database showed that little photometric information was available for them, not sufficient to assess their membership status. Therefore, this comparison supports the robustness of the membership selection based on GES spectroscopic parameters, and shows that this information is overall consistent with that provided by photometric data.

We then examined the sample of GES targets not classified as members based on GES spectroscopic criteria but with membership flag = 1 assigned in the CSI~2264 catalog. This sample amounts to 117 objects. A comparison between the spectroscopic information available from GES for these sources and their photometric properties prompted us to reject membership for 29 of these 117 objects, as they had weaker\footnote{For several objects in this group, membership had been assigned based mostly on their photometric data being consistent with the cluster locus on a CMD, and on their RV being consistent with the cluster distribution in the study of \citet{furesz06}. However, as shown in the same paper, the sample of cluster members selected upon their RV may be contaminated by field stars.} membership attribution from the CSI~2264 database and were classified as non-members based on their GES $\gamma$-index and/or EW(Li) parameters. On the other hand, this analysis allowed us to complement the spectroscopic member selection among GES targets by retrieving the other 88 objects. These comprise 65 sources with no membership information from GES data, three sources that had been discarded upon having a $\gamma$-index slightly larger than 1 (between 1 and 1.01) although exhibiting EW(Li) consistent with being members of the cluster, and another dozen of objects that had been discarded upon having EW(Li) lower than the threshold, even if larger than the average EW(Li) of field stars in the spectral range of interest (mainly early-M or K-type) by over 4\,$\sigma$ (i.e., EW(Li) $>$ 135~m\AA). For the latter two groups, we decided to consider these objects as members, despite the fact that they don't meet the membership requirements on GES parameters, because they appear to be borderline cases on those parameters, but exhibit strong indications of being PMS stars (e.g., strong X-ray emission, strong variability) in the CSI~2264 database. 

In principle, the selection of cluster members upon being strong X-ray emitters and variable sources might be affected by some degree of contamination by active field stars; these two aspects were illustrated respectively by \citet{flaccomio06} and \citet{venuti2015} for the PMS population of NGC~2264. However, as discussed in those papers, the expected rate of contaminants on these selection criteria is small ($\sim$4\% of the total sample). As a test, we examined the RV distribution of selected cluster members and flagged the objects (29) located $\geq$\,4\,$\sigma$ away from the center of the distribution and consistent with the RV properties of field stars in the NGC~2264 field. We then inspected the spatial distribution of these 29 sources across the region, to assess whether this is remarkably different from the spatial distribution of the 567 NGC~2264 spectroscopic members; however, the two groups exhibit similar spatial properties, and about 75\% of the 29 RV-outliers are projected onto the central areas of the cluster. We therefore conclude that, even if a small percentage of field contaminants are present in our member list, their impact for the statistical purposes of this study is negligible.

The final sample of objects probed in this study thus comprises 655 objects that were targeted during the GES observations of NGC~2264, and that were classified as probable cluster members based on either GES parameters or on complementary information available for these sources from the CSI~2264 database. A full list of these objects, along with the relevant parameters derived and/or discussed in this work, is provided in Table~\ref{tab:data}.

\subsection{Sample completeness}

\begin{figure*}
\centering
\includegraphics[width=\textwidth]{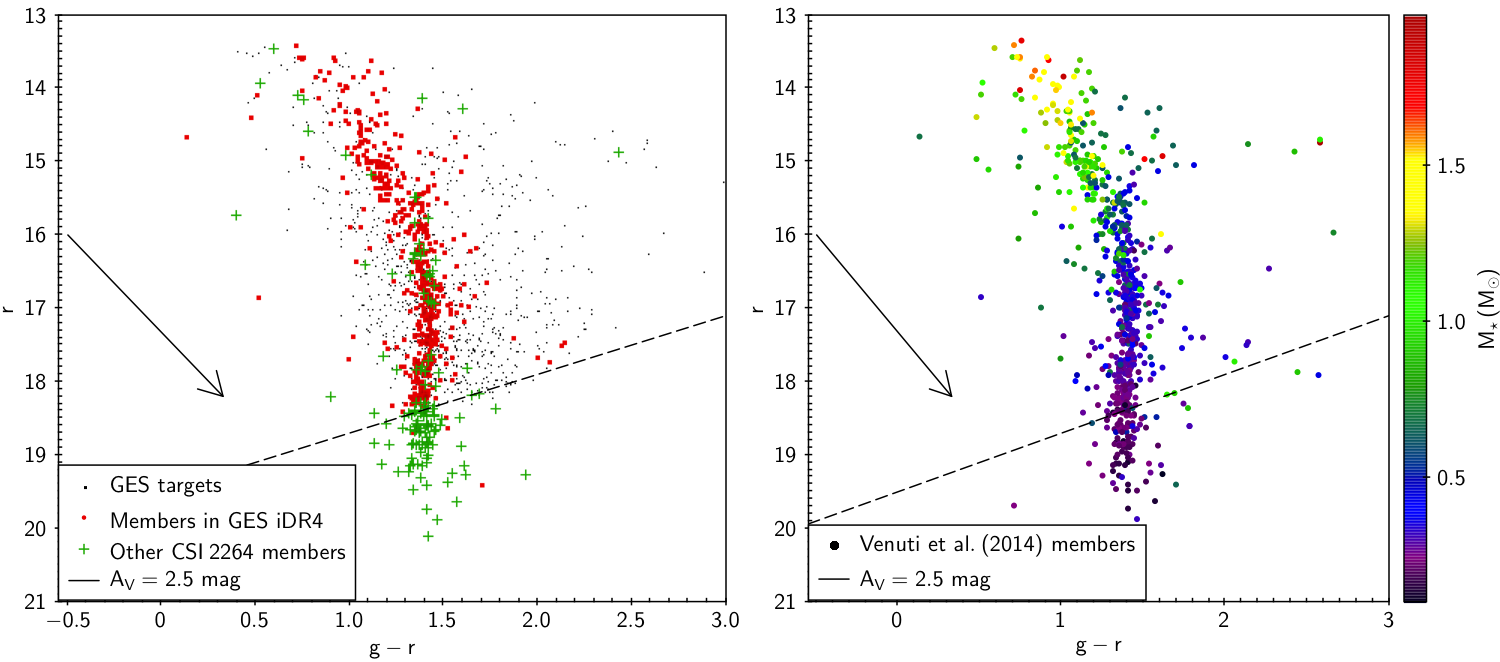}
\caption{{\it Left panel}: ($g-r$, $r$) CMD of the NGC~2264 region. The photometry used was acquired at the CFHT during the CSI~2264 campaign and is presented in \citet{venuti2014}. The different subgroups of objects shown on the diagram are: i) all objects in the NGC~2264 region with observations in the GES database (black dots); ii) GES targets classified as members as described in Sect.\,\ref{sec:memb_selection} (red dots); iii) additional confirmed or probable members, retrieved from the CSI~2264 catalog, that were not observed during the GES campaign (green crosses). The reddening vector is traced following the prescriptions of \citet{ADPS} in the SDSS photometric system. The black dashed line is traced by eye along the faint magnitude edge of the distribution of GES targets on the diagram to better visualize the magnitude limit of the GES observing run on NGC~2264. {\it Right panel}: same photometric diagram shown in the left panel, but for NGC~2264 cluster members investigated in \citet{venuti2014}. Colors are scaled according to stellar mass (derived in \citealp{venuti2014}, using PMS evolutionary tracks from the models of \citealp{siess2000}), as indicated in the side axis.}
\label{fig:gr_r_ges_csi}
\end{figure*}

To evaluate how representative the sample of members derived in Sect.\,\ref{sec:memb_selection} is with respect to the total PMS population of NGC~2264, we take as reference the master catalog of the NGC~2264 region issued by the CSI~2264 campaign \citep{cody2014}. This was built with the purpose to collect all sources, members and not, that were identified in the NGC~2264 field in the course of the many campaigns that have targeted this region in the literature, and is the most complete catalog of NGC~2264 available to date. We extracted from this catalog a full list of confirmed or likely cluster members of NGC~2264, and selected those that were not targeted during GES observations. In Fig.~\ref{fig:gr_r_ges_csi}, we plot the ($g-r$, $r$) CMD of the region\footnote{Figure~\ref{fig:gr_r_ges_csi} depicts, for illustration purposes, only objects that have $g,r,i$ photometry from the CFHT survey of the cluster \citep{venuti2014}. These amount to 73\% of the total sample of GES targets, 76\% of cluster members included in GESiDR4, and 32\% of the additional cluster members that appear in the CSI~2264 catalog of the region but are not included in the GES sample. However, we did explore a variety of photometric datasets to include a maximum of objects in our statistical evaluation of sample completeness, and could ascertain that the respective distributions observed on Fig.\,\ref{fig:gr_r_ges_csi} for the various subsamples are representative of the full samples (that is, irrespective of the photometric dataset considered, we consistently find that about 80\% of known cluster members with no data in GESiDR4 lie below the faint mag limit of the GESiDR4 sample).}, and compare the photometric properties of i) all GES targets in the NGC~2264 field, ii) cluster members among GES targets, and iii) additional cluster members not included among GES targets.

\begin{figure}
\resizebox{\hsize}{!}{\includegraphics{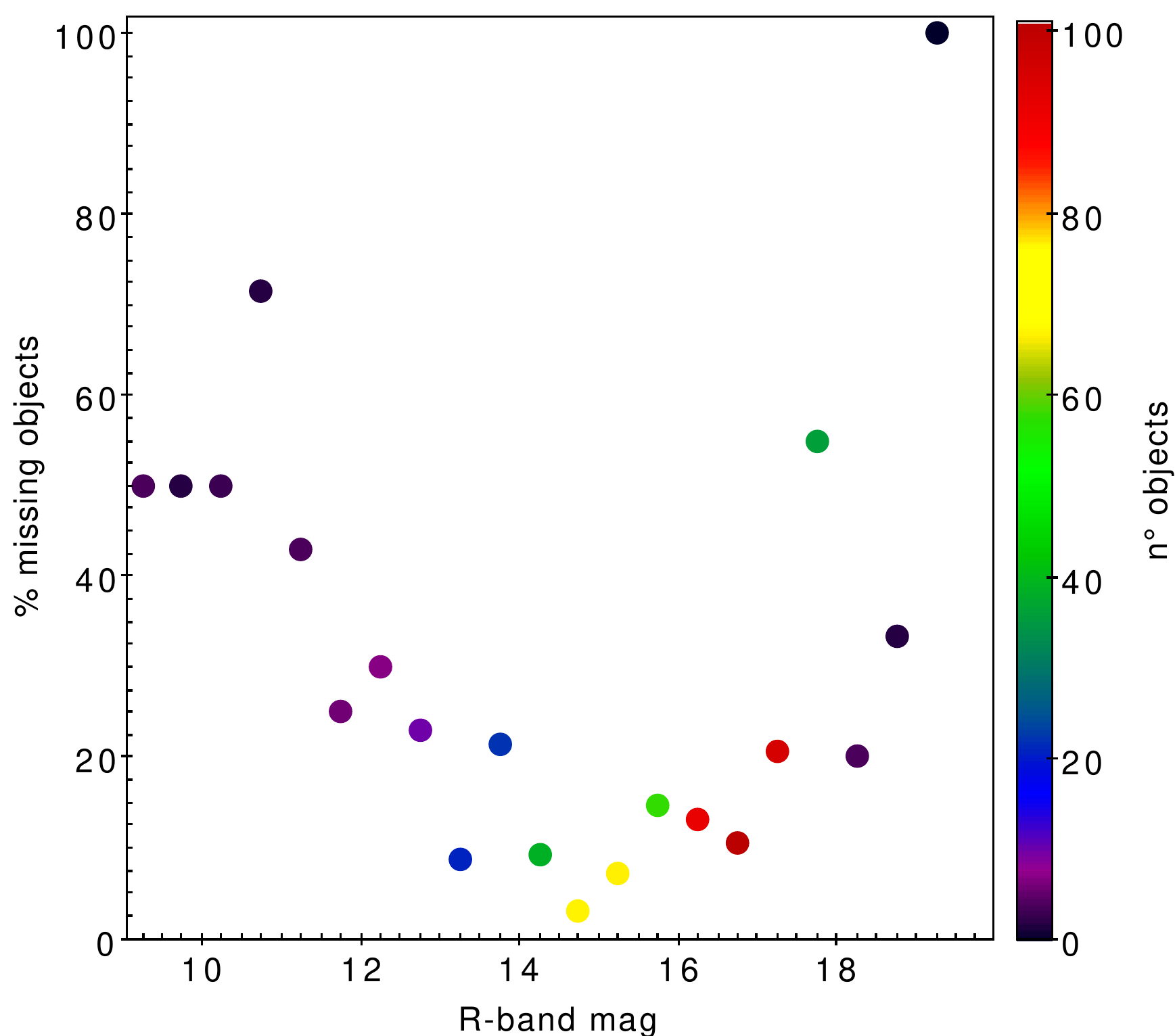}}
\caption{Fraction of missing cluster members in the GESiDR4 sample with respect to the full list of members gathered in the CSI~2264 database, as a function of R-band magnitude. The x-axis coordinate corresponds to the center of each mag bin used (see text). Colors are scaled according to the number of objects comprised in the GESiDR4 sample for each magnitude bin.}
\label{fig:GES_R_completeness}
\end{figure}

The CSI~2264 catalog contains about 800 sources listed as likely members of the cluster that are not included in the GES sample. However, about 80\% of them are fainter than 19~mag in the V-band \citep{sung08}, which corresponds to the limiting magnitude set for the NGC~2264 GES sample. This cut in magnitude occurs at a stellar mass of about 0.2~M$_\odot$ (Fig.\,\ref{fig:gr_r_ges_csi}, right panel). To appraise the potential impact of the remaining 20\% on the sample of members investigated here, we sorted GESiDR4 cluster members into R-mag bins of 0.5, and counted the number of additional members from the CSI~2264 list that would fall in each on these mag bins ($N_{CSI\,2264}(R_i)$). We then estimated, for each bin, the fraction of missing objects from the GESiDR4 sample, defined as 
\begin{equation}
\centering
\frac{N_{CSI\,2264}(R_i)}{N_{CSI\,2264}(R_i)+N_{GES\,iDR4}(R_i)}\,.
\end{equation}
The results of this computation are shown, as a function of R-band mag, in Fig.\,\ref{fig:GES_R_completeness}. Over 80\% of the cluster members included in the GESiDR4 sample have R-band magnitudes in the range 14 -- 17.5 mag, that roughly corresponds to a mass range 0.2--1.2~M$_\odot$. As can be seen, the majority of known cluster members with magnitudes in this range are encompassed by the GESiDR4 sample, with a completeness of around 90\% and fairly uniform across the magnitude range. The completeness level starts to decrease for magnitudes brighter than 13 in the R-band, where only a few tens of objects are available. We note here that, while the bulk of the GES observing run on NGC~2264 between V=12~mag and V=19~mag was performed with GIRAFFE, that allows targeting of up to 132 sources in a single pointing, spectra for sources brighter than V=12~mag (R$\sim$11.7~mag) could only be acquired with UVES, that enables observations of seven targets per pointing. While the GIRAFFE target selection was defined explicitly to be as complete as possible, UVES targets were carefully chosen to reduce the contamination with the main aim to use them to derive cluster metallicities and element abundances.

The completeness of our sample with respect to the known population of NGC~2264 is better than 90\% in the northern part of the cluster and in the immediate surroundings of the cluster cores, while it settles around 80\% in the more peripheral regions and in the southern part of the cluster, where the most embedded regions are located and stars are more absorbed and fainter (see later Sections). 

\subsection{Fraction of disks and accretors among the sample} \label{sec:disk_acc}

To separate NGC~2264 members in our sample according to their disk properties (which presumably reflect different evolutionary statuses), we followed \citet{cody2014}, who used near- and mid-infrared photometry from CSI~2264 ({\it Spitzer}/IRAC) and previous observations to classify IR~excess and IR~excess-free sources by measuring the slope of their spectral energy distribution (SED) between 2 and 24~$\mu$m. About 58\% of sources in our sample (377) do not exhibit signatures of a significant excess emission in the mid-IR, which suggests that they are devoid of material in the stellar surroundings; we therefore consider them as disk-free. Conversely, about 30\% of sources in our sample (197) exhibit significant IR excess, which indicates that they are still surrounded by a dusty disk. For the remaining 81 objects (12\% of our sample), we did not have IR information that would allow us to classify the sources as either disk-bearing or disk-free. 

In their study on the nature of the disk population of NGC~2264, \citet{teixeira2012} used the $\alpha_{IRAC}$ indicator, defined as the SED slope between 3.6 and 8.0~$\mu$m, to classify the evolutionary status of the sources with respect to their inner circumstellar material (see also \citealp{lada06alphaIRAC}). Following their scheme, the majority (88.6\%) of sources with no signatures of IR excess in our sample would be classified as ``naked photospheres'', while the remaining 11.4\% would fall in the ``anaemic disk'' category. Similarly, among objects with evidence of IR excess, the majority (78.6\%) have associated $\alpha_{IRAC}$ parameters that would qualify them as ``thick disks'', while smaller fractions appear to be in a more embedded (``flat spectrum'') or more evolved (transitional disks, anaemic disks) status. 

To identify objects with evidence of ongoing disk accretion, we primarily used information on the UV excess from CSI~2264 CFHT observations of the cluster \citep{venuti2014}. Indeed, the UV excess provides the most direct diagnostics of accretion onto the star, as it probes the emission in the Balmer continuum from the accretion shock generated by the impact of the column of accreting material onto the stellar surface. Measuring the luminosity of emission lines that originate in the accretion columns (among which the most prominent is the H$\alpha$ line) also provides a good tracer of ongoing accretion processes, as illustrated in several studies \citep[e.g.,][]{alcala2014}. Based on \citeauthor{venuti2014}'s (\citeyear{venuti2014}) results, we could identify 109 young stars with evidence of accretion in our sample from their measured UV excess, while for another 326 the UV data from CFHT do not reveal any discernible signature of ongoing accretion. 

For sources that lack UV excess data, we examined, in order of preference, the H$\alpha$,\,$r$,\,$i$ photometry available for the cluster from the IPHAS survey \citep{iphas, barentsen2014}, or information on the H$\alpha$ emission from GES data \citep{lanzafame2015}. We used the photometric H$\alpha$ information from IPHAS in preference to the GES spectroscopy, because the latter might be affected by subtraction of the nebular background emission (e.g., \citealp{bonito2013}; see \citealp{lanzafame2015}, and also Bonito et al., in preparation, regarding the specific case of NGC~2264 in GESiDR4). While CCD photometry allows local background subtraction in an aperture around each extracted source, this approach cannot be applied to spectroscopic surveys. As briefly addressed in \citet{jeffries2014}, the sky subtraction issue for GES GIRAFFE data is handled as follows: a number of instrument fibers, distributed across the field, are assigned to the sky, and the measured spectra are combined to derive the median background emission spectrum to be subtracted from the observed stellar spectra (see also Lewis et al., in preparation). However, the nebular emission in the NGC~2264 region is known to be spatially variable, which renders this procedure uncertain and hampers any quantitative measurements. For this reason, we expect photometric H$\alpha$ emission data to be potentially less affected\footnote{This might not be true in cases where the local background exhibits rapid spatial variations on very small scales, as illustrated in \citet{kalari2015}.} by the local nebular emission component than spectroscopic H$\alpha$ line parameters. 

\begin{figure}
\resizebox{\hsize}{!}{\includegraphics{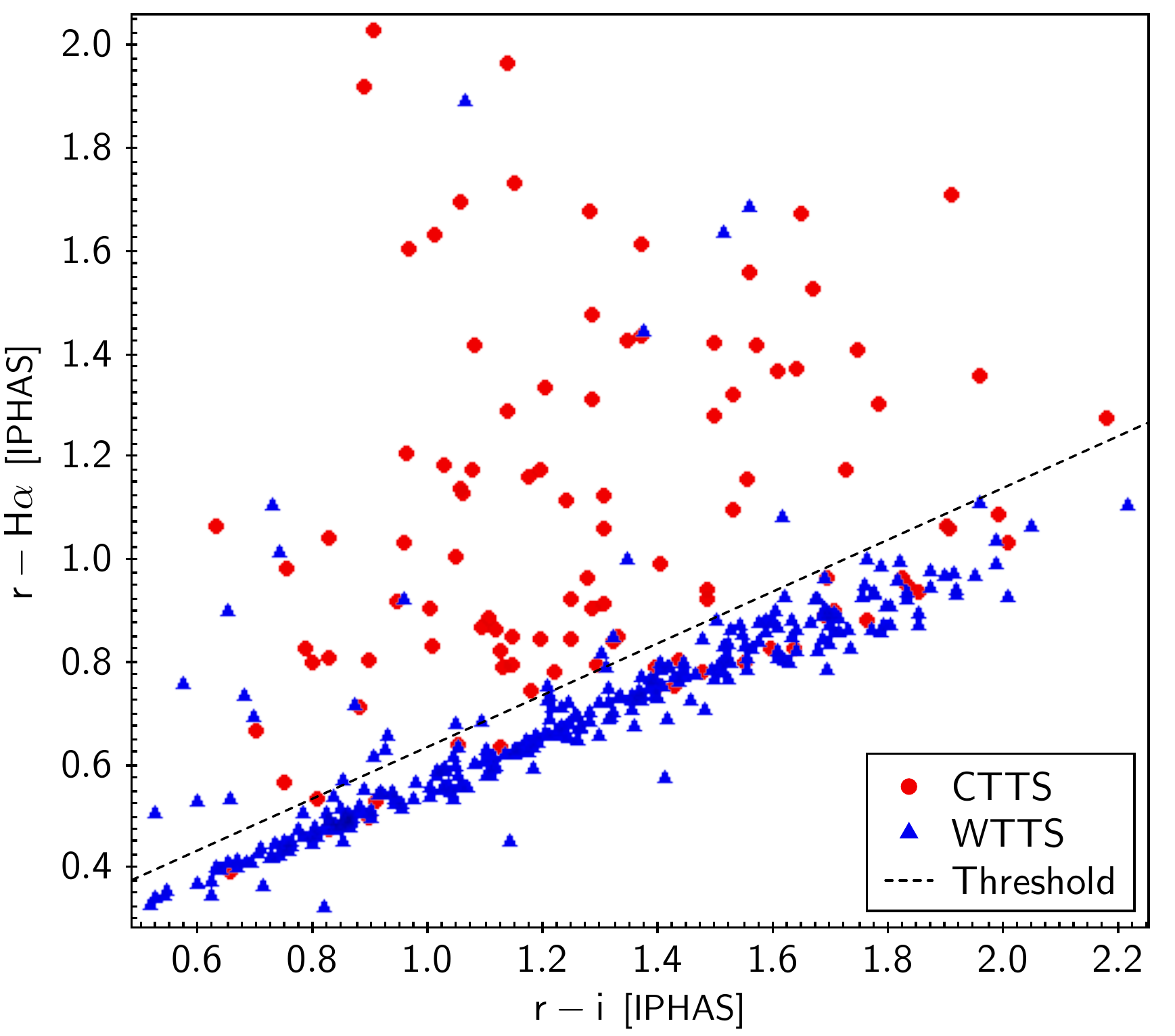}}
\caption{$r-H\alpha$ vs. $r-i$ color-color diagram using IPHAS photometry for NGC~2264 members in our sample classified as either accreting (CTTS, red dots) or non-accreting (WTTS, blue triangles) based on the UV excess measured for these sources in \citet{venuti2014} from CSI~2264 CFHT photometry. The black dashed line is an empirical threshold traced following the locus of WTTS to separate the regions dominated by accreting vs. non-accreting objects; this threshold follows closely the unreddened MS track tabulated by \citet{iphas} on the $r-H\alpha$ vs. $r-i$ diagram.}
\label{fig:IPHAS_accreting}
\end{figure}

To select likely accreting members from IPHAS data, we used the samples of accreting and non-accreting objects selected from the UV excess diagnostics to identify the region of the ($r-i$, $r-H\alpha$) diagram that is dominated by accretion \citep[e.g.,][]{barentsen2011}, as shown in Fig.\,\ref{fig:IPHAS_accreting}.
We then selected all cluster members in our sample with IPHAS data but not CFHT UV data, plotted them onto the ($r-i$, $r-H\alpha$) diagram, and classified as possible accreting sources all those that fall above the threshold used to separate the accretion-dominated region of the diagram in Fig.\,\ref{fig:IPHAS_accreting}. These amount to 37 objects, that add to the 109 accreting sources identified based on the UV excess diagnostics. To complete our selection of accreting sources in our sample, we finally examined the spectroscopic H$\alpha$ parameters from GES for objects that lack both UV excess information from CSI~2264 and photometric H$\alpha$ information from IPHAS. To be conservative, we adopted a threshold of 270~km/s on the W10\%(H$\alpha$) parameter and a $T_{\rm eff}$--dependent threshold on EW(H$\alpha$), following \citet{white_basri2003}, and retained only objects with evidence of ongoing accretion on both criteria, unless only one of the two measures is available for a given source. This led to the identification of ten additional members with signatures of accretion in our sample. We cross-checked the original spectra of the sources, the background spectra used for the correction, and the background-subtracted spectra used for scientific analysis for these ten objects; this test allowed us to ascertain that, although the procedure of background subtraction might somewhat affect the intensity of the H$\alpha$ line, it never alters the original shape of the line for the cases of interest. For the purposes of this work, we can therefore be confident that the EW(H$\alpha$) measured on these spectra is truly indicative of accretion.

Finally, the UV excess diagnostics to trace ongoing accretion may be ineffective in the case of young star-disk systems seen at high inclinations (i.e., where the disk is seen almost edge-on; see, e.g., \citealp{bouvier2007a}, \citealp{fonseca2014}, \citealp{mcginnis2015}). In this geometric configuration, the accretion spots at the stellar surface, responsible for the UV excess emission, may be concealed from view by structures in the inner disk. On the other hand, the H$\alpha$ emission profile detected for these sources would reveal if they are actively accreting \citep[e.g.,][]{sousa2016}. We therefore re-examined the sample of 326 sources with no evidence of UV excess from CFHT data, and flagged those that would qualify as accreting based on the H$\alpha$ emission parameters from IPHAS and GES data (see the blue triangles above the threshold line in Fig.\,\ref{fig:IPHAS_accreting}). A visual inspection of their H$\alpha$ emission profiles (cf. \citealp{reipurth1996}, \citealp{kurosawa2006}) and their {\it CoRoT} light curves (cf. \citealp{cody2014}) allowed us to corroborate the accreting status for 31 stars in this group.

The final list of accreting objects thus comprises 187 sources (28.5\% of our sample), while another 409 (62.4\%) have either UV data or photometric H$\alpha$ data that indicate no significant accretion activity. To estimate the fraction of contaminants that may affect our sample due to the variety of accretion diagnostics used, we selected putative accretors based on, in turn, photometric H$\alpha$ from IPHAS or spectroscopic H$\alpha$ from GES, and checked in how many cases, if an UV excess measurement from CSI~2264 data is available, this would also qualify the object as accreting. The comparison showed that about 76\% of objects selected as accreting on the ($r-i$, $r-H\alpha$) diagram would also be classified as accretors based on their UV excess, while the remaining 24\% do not exhibit any significant UV excess. These fractions become 59\% and 41\%, respectively, when we consider the sample of spectroscopic H$\alpha$ accretors with respect to the UV excess diagnostics. This result strengthens our assumption that, in the absence of UV excess information, IPHAS photometric H$\alpha$ provides a more robust diagnostics of accretion than GES spectroscopic H$\alpha$. Since our final sample of accreting objects contains 78 sources selected based on photometric or spectroscopic H$\alpha$ emission, we can estimate that $\sim$8\% (about a dozen objects) are potential contaminants. This analysis does not take into account that accretion is intrinsically a variable process, and hence some objects may exhibit signatures of ongoing accretion at certain epochs and not in others. However, the study of accretion variability in NGC~2264 presented in \citet{venuti2014} showed that, at least on timescales of a few weeks, which provide the dominant component of variability in young stars on timescales of up to several years \citep{venuti2015}, only a small fraction of disk-bearing sources ($\sim$8\%) exhibit a fluctuating behavior around the accretion detection threshold. This suggests that our strategy to identify accreting sources in our sample may lead to omitting a few ($\sim$ten) objects; for the statistical purposes of our work, we assume this component to be negligible.

For 530/655\footnote{For the remaining 125 objects, no classification either on the disk diagnostics or on the accretion diagnostics, or both, is available, due to missing or ambiguous data. In these cases, a blank field is reported in the ninth and/or tenth column of Table~\ref{tab:data}. We examined the distribution of these sources on CMD and R.A.--Dec diagrams with respect to the objects for which a classification on disk and accretion properties is available, but did not detect any obvious bias in magnitude or spatial location within the cloud.} objects in our sample, we had sufficient data to derive a classification in both the disk-bearing/disk-free (Class~II/III; IR) and in the accreting/non-accreting (UV, H$\alpha$) categories (ninth and tenth column, respectively, of Table~\ref{tab:data}). Of these 530 sources, 190 exhibit disks; among these, 74.2\% are accreting, while the rest are classified as non-accreting. Among disk-free sources with also information on accretion (340/530), the vast majority (95.6\%) are classified as non-accreting according to our criteria, while a few objects exhibit signatures consistent with ongoing accretion. The latter group could correspond to more evolved disks, with no IR excess emission detectable in the shorter-$\lambda$ IRAC channels, where some accretion activity is still ongoing \citep[e.g.,][]{sicilia_aguilar2006}, or to objects with enhanced cromospheric activity; at any rate, these few cases of more uncertain classification do not impact the statistical soundness of our investigation. 

In the following, we refer to the group of objects with IR excess signatures in our sample as ``Class~II objects'', and to those with no IR excess signatures as ``Class~III objects'' \citep[cf.][]{lada1987}. We use instead the nomenclature ``classical T~Tauri stars'' (CTTS) to refer to accreting sources in our sample, and ``weak-lined T~Tauri stars'' (WTTS) to refer to non-accreting sources in our sample \citep[cf.][]{herbig_bell1988}.

\section{Results} \label{sec:results}

\subsection{A$_V$ estimates for cluster members} \label{sec:Av}

\begin{figure}
\resizebox{\hsize}{!}{\includegraphics{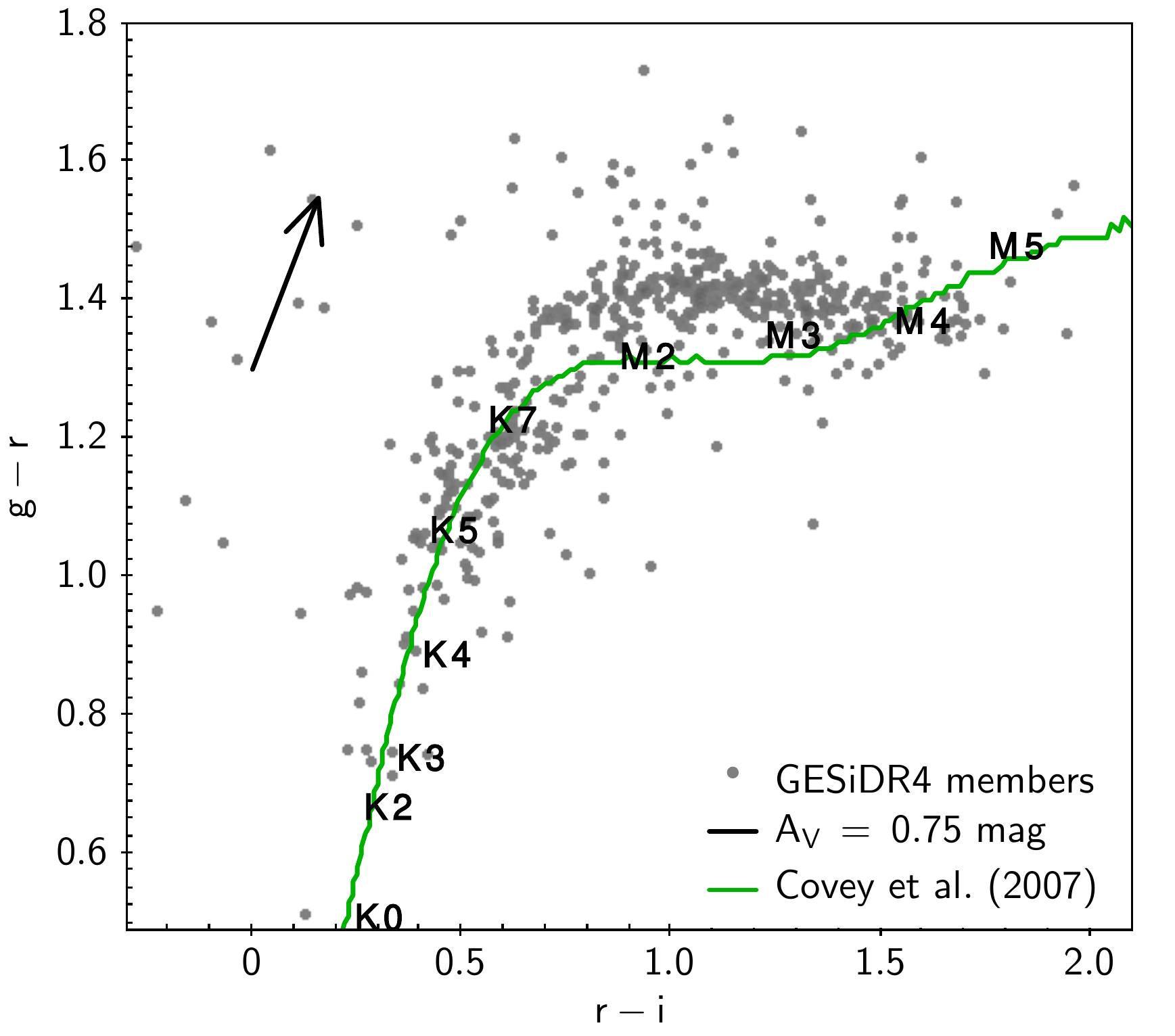}}
\caption{Distribution of NGC~2264 members on a $g-r$ vs. $r-i$ diagram, built using the SDSS-calibrated photometry obtained for the cluster with CFHT/MegaCam \citep{venuti2014}. The green line shows the empirical, zero-reddening SpT--color stellar locus derived in the statistical study of \citet{covey07}, recalibrated onto CFHT photometry as in \citet{venuti2014}. The reddening vector is traced following \citet{ADPS}.}
\label{fig:Av_GES_gri}
\end{figure}

\begin{figure*}
\centering
\includegraphics[width=\textwidth]{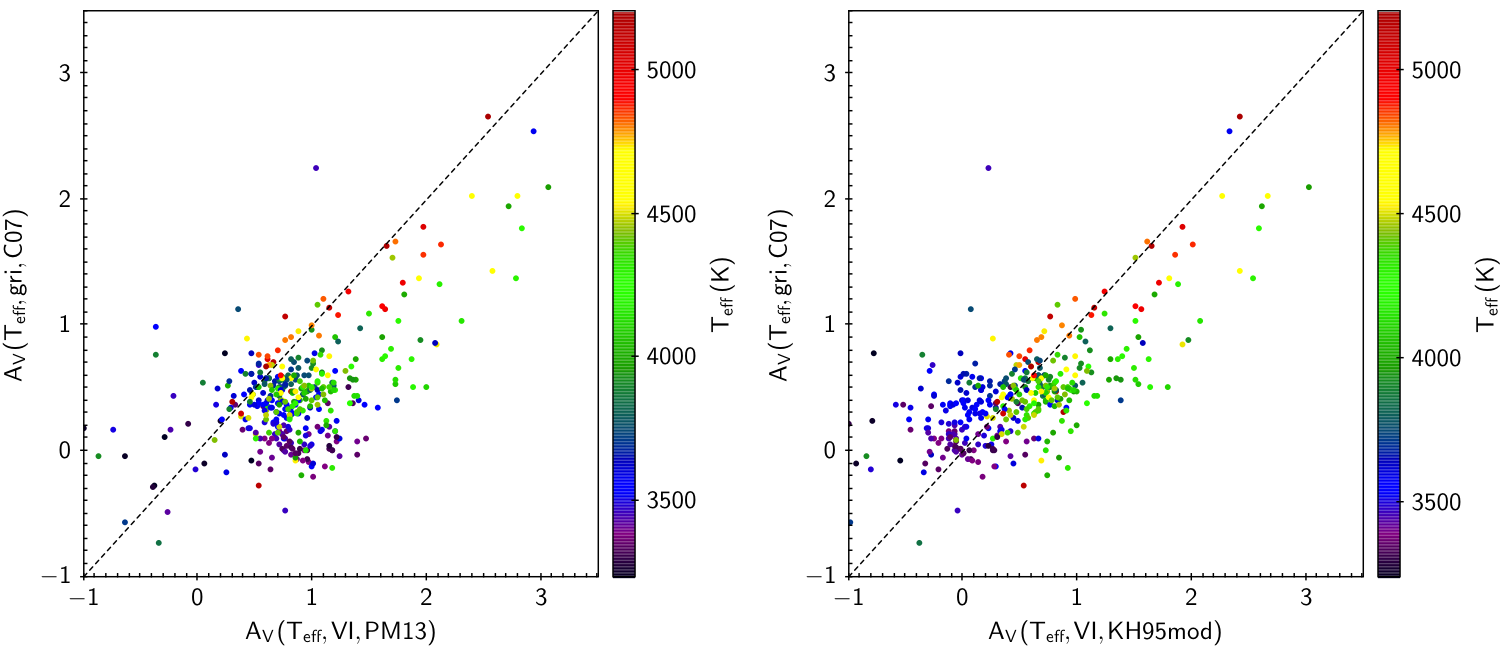}
\caption{{\it Left panel}: comparison between the A$_V$ estimates derived for cluster members following methods 1. (y-axis) and 2. (x-axis) of Sect.\,\ref{sec:Av}. {\it Right panel}: comparison between the A$_V$ estimates derived following methods 1. (y-axis) and 3. (x-axis) of Sect.\,\ref{sec:Av}. Each dot corresponds to an object; colors are scaled according to the objects' $T_{\rm eff}$, as shown on the side axes of the diagrams. The dashed lines trace the equality line on the diagrams.}
\label{fig:Av_comp}
\end{figure*}

To derive an estimate of extinction parameters for individual objects and map reddening across NGC~2264, we followed three different approaches. These consisted in combining the effective temperatures $T_{\rm eff}$ provided by GES with three different photometric datasets and reference sequences for the photospheric colors of young stars; a comparison of the sets of A$_V$ derived from each method allowed us to select the most suitable derivation for our analysis. We chose to refer our A$_V$ derivation to optical photometry of the cluster, as near-infrared (J,H,K) photometry, although covering our sample more extensively, may be affected by thermal emission from the disk that would introduce an additional source of scatter in our results. 

The three independent methods we used to determine the A$_V$ values are enumerated below.

\begin{enumerate}
\item \emph{A$_V$\,($T_{\rm eff}$,\,gri,\,C07)}. In a first derivation, we used the CFHT $g,r,i$ photometry available from CSI~2264, calibrated onto the SDSS photometric system as described in \citet{venuti2014}; we took as reference the stellar locus on the ($r-i$, $g-r$) diagram tabulated empirically as a function of spectral type (SpT) in the study of \citet{covey07}, based on a sample of several hundred thousand objects detected across the Galaxy by the SDSS \citep{SDSSDR2}. This approach allows us to encompass 507/655 objects in our sample, while for the remaining sources no complete or good quality $g,r,i$ photometry is available. The observed colors of the 507 objects and the expected color sequence as a function of SpT are shown in Fig.\,\ref{fig:Av_GES_gri}. We used the $T_{\rm eff}$ associated with each object to determine the expected location of the corresponding object on the reference color sequence in the absence of extinction; we then computed A$_V$($g-r$) and A$_V$($r-i$) by measuring the displacement between the observed colors and the expected colors along the reddening vector for that source. When the two estimates were consistent with each other, we adopted an average between the two as the best A$_V$ estimate; when the two estimates were discrepant, we favored the A$_V$ estimate inferred from the $g-r$ color, as this was built from simultaneous observations \citep{venuti2014} and is more sensitive than $r-i$ to extinction and to variations in SpT across most of the spectral range of interest (Fig.\,\ref{fig:Av_GES_gri}).
\item \emph{A$_V$\,($T_{\rm eff}$,\,VI,\,PM13)}. Our second approach consisted in combining V,\,I photometry from literature studies \citep{sung08, rebull2002, lamm04} and \citeauthor{pecaut2013}'s (\citeyear{pecaut2013}) $T_{\rm eff}$\,--\,color scale for PMS objects. This approach could be extended to 482/655 objects in our sample. A$_V$ were estimated by computing the color excess E(V-I) for each object and assuming a standard reddening law.
\item \emph{A$_V$\,($T_{\rm eff}$,\,VI,\,KH95mod)}. Finally, we derived a third set of A$_V$ estimates by combining V,\,I photometry with the $T_{\rm eff}$ -- color sequence for dwarfs tabulated in \citet{kenyon1995}, modified following \citet{stauffer1998} for M-type stars. The methodology applied in this case is the same as that presented in point 2.
\end{enumerate} 

A comparison between the different sets of A$_V$ obtained is illustrated in Fig.\,\ref{fig:Av_comp}. As can be seen from the diagrams, the adoption of various $T_{\rm eff}$--color scales translates into A$_V$ estimates even significantly different from each other. In particular, method 2. yields values of A$_V$ that are systematically larger than those inferred from methods 1. and 3., albeit with a $T_{\rm eff}$--dependent specific offset (largest for M-type stars and smallest for early-K stars). An average offset of about 0.4~mag can be measured between the two sets of results from methods 1. and 2. (left panel of Fig.\,\ref{fig:Av_comp}); in addition, the cloud of points on the diagram also exhibits an internal dispersion of about 0.4~mag. The extent of this scatter (see also \citealp{cauley2012} and their Fig.~4) does not change appreciably if we retain only disk-free sources for the comparison. A $T_{\rm eff}$--dependence also appears in the offset between the A$_V$ estimates from methods 1. and 3. (right panel of Fig.\,\ref{fig:Av_comp}). In this case, there is no systematic displacement of one sequence below or above the other on the diagram: method 1. yields A$_V$ values typically larger than method 3. for M-type stars, and smaller than method 3. for K-type stars, with a resulting average offset of $\sim$0.06~mag and a dispersion of $\sim$0.4~mag. 

The derived A$_V$\,($T_{\rm eff}$,\,gri,\,C07) estimates exhibit some $T_{\rm eff}$ dependence for $T_{\rm eff} \leq 3500$~K (SpT of M2 and later): these objects appear on average less extincted than earlier-type stars. This behavior may reflect the mismatch between the empirical cluster locus and the reference color sequence for late-type stars that can be seen in Fig.\,\ref{fig:Av_GES_gri}. A similar $T_{\rm eff}$ dependence can be observed in the derived set of A$_V$\,($T_{\rm eff}$,\,VI,\,KH95mod): SpT\,$\geq$\,M1 objects tend to be associated with lower extinction values than K-type objects. Conversely, A$_V$\,($T_{\rm eff}$,\,VI,\,PM13) estimates appear to be roughly $T_{\rm eff}$--independent; however, the median A$_V$ derived from method 2. for objects later than M2 is 0.77~mag, and this is inconsistent with the photometric properties of M-type cluster members in Fig.\,\ref{fig:Av_GES_gri}. Indeed, if the lower envelope of the M-type color locus on the ($r-i$, $g-r$) diagram corresponds to non-reddened objects, an empirical upper limit to the A$_V$ of the bulk of M-type cluster members is provided by the width of their color locus along the reddening vector in Fig.\,\ref{fig:Av_GES_gri}; this amounts to $\sim$0.35~mag.

Given the significant scatter in values between two different A$_V$ calculations for the same sample of objects, we resolved to not combine results from different methods, but to select and adopt only one set of values, and restrict our analysis to cluster members with A$_V$ estimates from that method whenever dereddened photometry is needed. Since method 1. provided A$_V$ for a slightly larger sample of objects than methods 2. or 3., and makes use of a single and empiric reference color sequence, we ultimately decided to adopt A$_V$\,($T_{\rm eff}$,\,gri,\,C07). Arithmetic uncertainties on the derived values of A$_V$ follow from the uncertainties on the observed colors and on the expected colors; these, in turn, result from photometric errors as well as uncertainties on $T_{\rm eff}$ (that determines the expected location of a given source on the reference color sequence). Adopting an average $T_{\rm eff}$ uncertainty of 3\%, photometric errors from \citet{covey07} for the expected colors and from \citet{venuti2014} for the observed colors, and treating each source of error as independent from the others, we derived a typical uncertainty of 0.2~mag on our A$_V$ estimates. The actual uncertainty on the individual A$_V$ might be larger than this arithmetic estimate, as suggested by the amount of scatter between different A$_V$ estimates for the same sources in Fig.\,\ref{fig:Av_comp}.

\begin{figure}
\resizebox{\hsize}{!}{\includegraphics{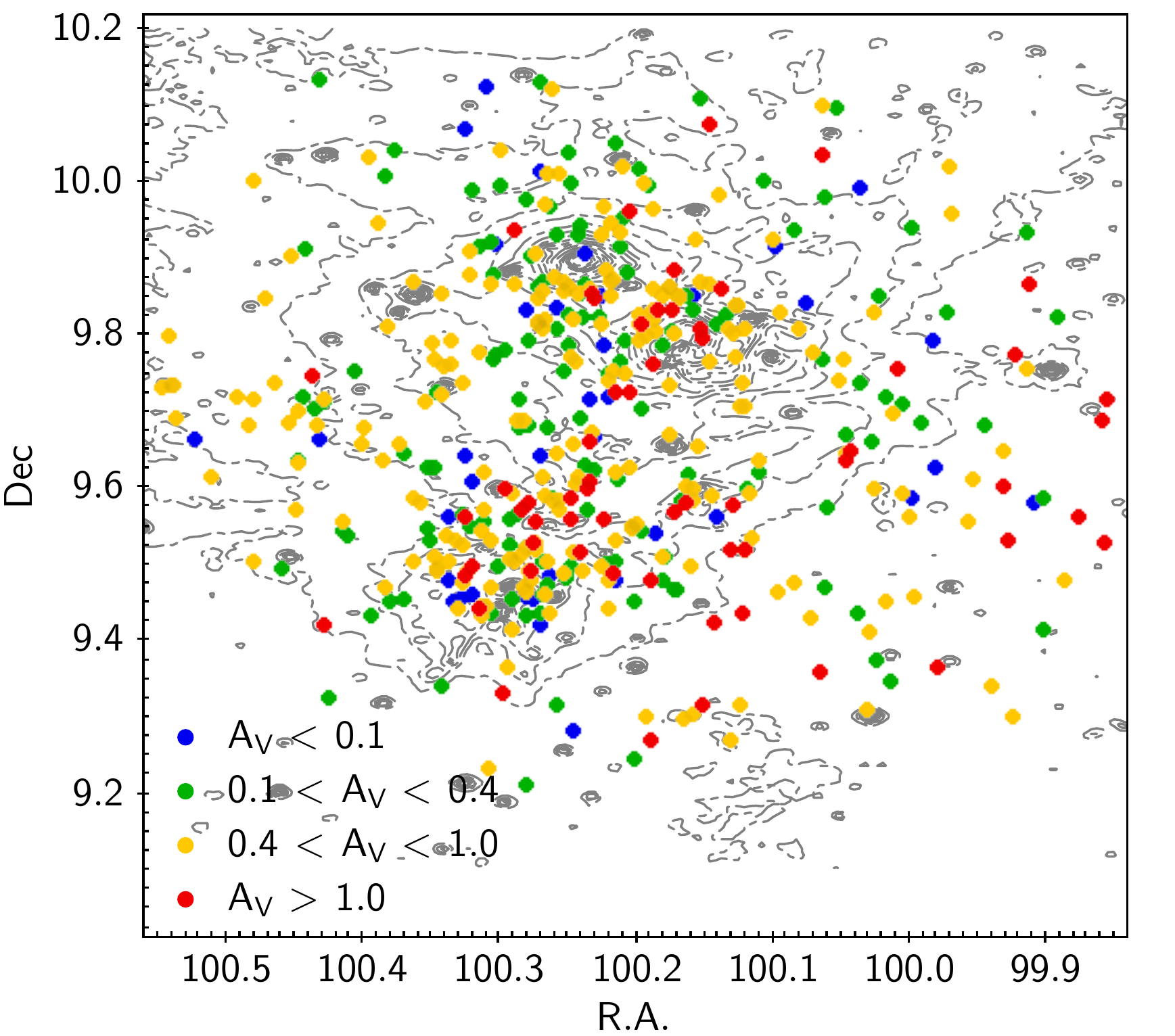}}
\caption{Spatial distribution of NGC~2264 cluster members investigated in this study as a function of their derived A$_V$. Four A$_V$ groups are distinguished on the diagram: A$_V$$<$0.1~mag (blue dots), 0.1$<$A$_V$$<$0.4~mag (green); 0.4$<$A$_V$$<$1.0~mag (yellow), and A$_V$$>$1.0~mag (red). Isocontours drawn in gray were extracted from a DSS colored optical image of the region using the Aladin Sky Atlas \citep{ALADIN}.}
\label{fig:Av_radec}
\end{figure}

The typical A$_V$ associated with cluster members appears to be small: the median A$_V$ derived in our sample is 0.44 mag, consistent with earlier findings (see \citealp{dahm08} and references therein). A $\sigma$ dispersion of 0.3~mag, derived applying an iterative 3\,$\sigma$--clipping procedure, is measured around this value. A few tens of objects have derived A$_V$$<$0; while for a part of them the actual value is consistent with zero (non-extincted object) within the dispersion, for others the significantly negative value of A$_V$ likely reflects the presence of strong disk accretion or scattered light from the disk plane that determine the objects to appear bluer than they actually are \citep[e.g.,][]{guarcello2010b}, or perhaps photometric issues that result in an incorrect positioning of the source on the color-color diagram in Fig.\,\ref{fig:Av_GES_gri}. When comparing the distributions in A$_V$ obtained for Class~II and Class~III sources separately, a somewhat higher dispersion (0.5~mag) is detected across the first group relative to the second. However, the average statistical properties of the A$_V$ distribution is not significantly affected by the presence or absence of circumstellar disks, and the median values measured for the two cases (0.43~mag for Class~II objects and 0.45~mag for Class~III objects) are well consistent with each other within the associated uncertainties. The spatial distribution of cluster members as a function of their A$_V$ is shown in Fig.\,\ref{fig:Av_radec}. As suggested by the distribution of colors in the central part of the diagram, more heavily extincted objects tend to be clustered around the innermost regions of NGC~2264, while little extincted objects tend to distribute more evenly across the cloud.

\subsection{The HR diagram of NGC~2264} \label{sec:HR}

To explore the properties of the NGC~2264 population and estimate mass and age of individual members, we placed the objects on a HR diagram. As mentioned in Sect.\,\ref{sec:introduction}, while this would be problematic for embedded protostars \citep{stahler_palla}, the HR diagram is a suitable tool to trace the evolution of PMS objects (on which this study is focused), although the apparent positions of CTTS might be affected by the presence of disks and ongoing accretion. 

Bolometric luminosities L$_{bol}$ were determined using the dereddened $i$-band photometry; we opted to use this filter ($\lambda_{eff}$~=~744~nm, slightly blueward of the Johnson-Cousins I-band; \citealp{bessell2005}) as it is substantially unaffected by either accretion luminosity or thermal emission from the disk, and the impact of extinction is more limited at these wavelengths as compared to the V or R bands \citep{kenyon1990}. To associate bolometric corrections BC$_i$ to the $i$-band magnitudes, we used the $T_{\rm eff}$ -- dependent scale tabulated by \citet{girardi2008}\footnote{http://stev.oapd.inaf.it/dustyAGB07/} in the SDSS filters, for $\left[M/H\right]=0$ and $\log{g}=4.5\footnote{This value was chosen by computing the median $\log{g}$ for objects in our sample with gravity parameters released in GESiDR4. An estimate of $\log{g}$ from GES is available for only 47 NGC~2264 members with an estimate of A$_V$; the median $\log{g}$ measured across this subgroup is 4.47.}$. We interpolated between two successive points in \citeauthor{girardi2008}'s (\citeyear{girardi2008}) table to associate BC$_i$ estimates with intermediate values of $T_{\rm eff}$. To derive an estimate of the typical uncertainty on L$_{bol}$ values, we applied stardard error propagation rules accounting for uncertainties on $i$-band photometry, A$_V$, and BC$_i$ (in turn dependent on the uncertainty on $T_{\rm eff}$); this procedure provided $\sigma_{L_{bol}}$/L$_{bol}$\,$\sim$\,0.13.

\begin{figure}
\resizebox{\hsize}{!}{\includegraphics{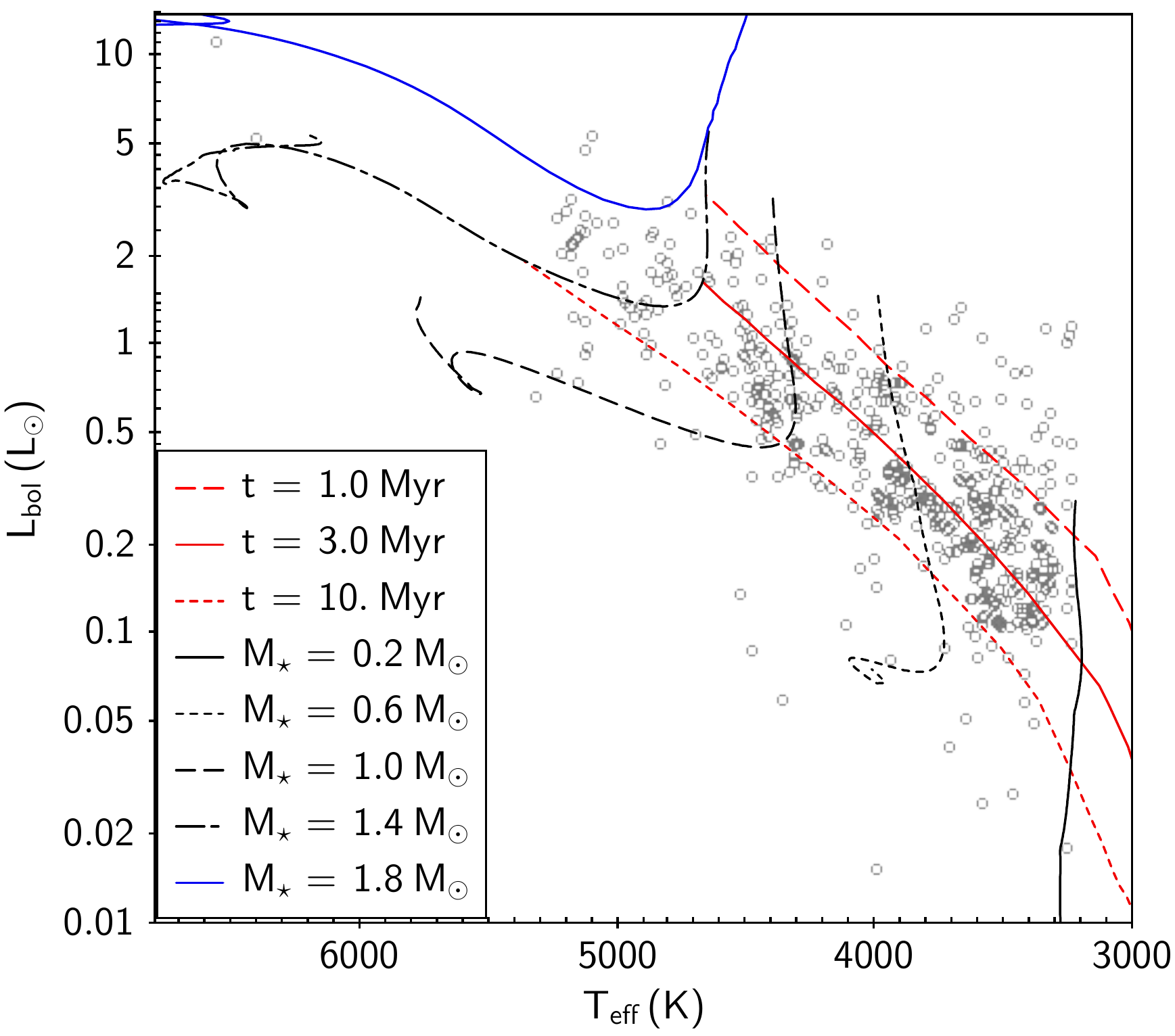}}
\caption{HR diagram for NGC~2264 members investigated in this study (gray circles). Isochrones at 1, 3 and 10~Myr, traced in red on the diagram, are extracted from \citeauthor{baraffe2015}'s (\citeyear{baraffe2015}) models, as well as mass tracks in the 0.2--1.4~M$_\odot$ mass range (black curves); mass tracks from \citeauthor{siess2000}'s (\citeyear{siess2000}) PMS evolutionary models are adopted for objects that lie above \citeauthor{baraffe2015}'s $M_\star = 1.4~M_\odot$ track, that is, the highest mass included in the PMS model tracks published by \citet{baraffe2015}. The mass track at $M_\star = 1.8~M_\odot$ from \citet{siess2000} is shown in blue as an example.}
\label{fig:HR_models}
\end{figure}

The resulting HR diagram for the cluster is shown in Fig.\,\ref{fig:HR_models}.
A distinctive feature of this diagram is the significant vertical spread observed at any given $T_{\rm eff}$. The most straightforward interpretation of this feature is that the population of the cluster exhibits an internal age spread of several Myr; however, several studies have cautioned against inferring age spreads from apparent luminosity spreads on HR diagrams, arguing that external effects such as the individual accretion history of cluster members \citep{baraffe2009}, the presence of surface spots \citep{gullysantiago2017}, unresolved binarity, and observational uncertainties \citep[e.g.,][]{soderblom2014} may mimic a spread of several Myr among perfectly coeval stars. An extended discussion on the nature of the vertical spread observed for NGC~2264 members on the HR diagram is provided in Sect.\,\ref{sec:age_spread}.

To associate an estimate of mass and age with individual stars, we used the grid of PMS evolutionary models recently developed by \citet{baraffe2015}, and interpolated the position of each source on the HR diagram between the two closest mass tracks and isochrones available. \citeauthor{baraffe2015}'s (\citeyear{baraffe2015}) mass tracks cover the $M_\star$ range from 0.07~M$_\odot$ to 1.4~M$_\odot$. To derive a mass estimate for the few tens of objects that fall in the $M_\star > 1.4~M_\odot$ region of the HR diagram in Fig.\,\ref{fig:HR_models}, we adopted mass tracks from \citeauthor{siess2000}'s (\citeyear{siess2000}) models. The orientation of these mass tracks on the HR diagram does not match perfectly that of \citeauthor{baraffe2015}'s grid, as can be noticed on Fig.\,\ref{fig:HR_models}; however, this mismatch does not impact the relative scale in mass that we derive across our sample, since the two sets of models are applied to two separate mass regimes, and both sets of models are concordant in assigning $M_\star \leq 1.4~M_\odot$ to objects in the region of the HR diagram covered by \citeauthor{baraffe2015}'s model grid, and $M_\star \geq 1.4~M_\odot$ to objects falling above this region on the HR diagram in Fig.\,\ref{fig:HR_models}. Moreover, recent studies \citep{alcala2017, frasca2017} have shown that the mass estimates derived using PMS model tracks from \citet{baraffe2015} are in very good agreement with those derived using \citeauthor{siess2000}'s (\citeyear{siess2000}) PMS tracks in the region where the two sets of models overlap. On the other hand, objects at $M_\star < 1.4~M_\odot$ and those at $M_\star > 1.4~M_\odot$ virtually distribute along the same isochrones in Fig.\,\ref{fig:HR_models}; therefore, to avoid any bias that may ensue from combining isochrones associated with two different sets of models (which exhibit some small shift with respect to each other for a nominal age), we decided to restrict the investigation of isochronal ages only to objects encompassed by \citeauthor{baraffe2015}'s (\citeyear{baraffe2015}) model grid.

Stellar parameters (mass and age) were determined for 443/507 (87.4\%) of the objects for which we could derive an estimate of A$_V$ as discussed in Sect.\,\ref{sec:Av}. Among the 64 objects with $T_{\rm eff}$ and L$_{bol}$ estimates but with no estimate of mass and/or age inferred from Fig.\,\ref{fig:HR_models}, 19 fall above the 0.5~Myr isochrone, that corresponds to the lowest age track included in \citeauthor{baraffe2015}'s (\citeyear{baraffe2015}) models; six (likely edge-on disks, or sources affected by photometric issues) fall below the region of the HR diagram covered by PMS mass tracks; 39 fall above the $M_\star = 1.4~M_\odot$ track (for these cases, we estimated the mass from the models of \citealp{siess2000} but no estimate of age, as explained earlier). Derived masses range from 0.2~$M_\odot$ to 1.8~$M_\odot$; the median isochronal age measured across our sample\footnote{This estimate only takes into account objects for which an estimate of age from \citeauthor{baraffe2015}'s (\citeyear{baraffe2015}) isochrones could be derived. It does not include, for instance, the about twenty objects that appear to be younger than 0.5~Myr based on their position on the HR diagram in Fig.\,\ref{fig:HR_models}.} is 3.6~Myr ($\log{yr} = 6.55$), with a dispersion of 0.35~dex around this value. We note that slightly different results on L$_{bol}$ and age would be obtained across our sample if we adopted the A$_V$ values derived from the methods 2. or 3. listed in Sect.\,\ref{sec:Av}. L$_{bol}$ estimates systematically larger by a factor of $\sim$1.35 would be derived using the A$_V$\,($T_{\rm eff}$,\,VI,\,PM13) values; this would determine the cluster locus on the HR diagram to be systematically shifted toward younger ages, settling around an average value of $\sim$2~Myr. Using the A$_V$\,($T_{\rm eff}$,\,VI,\,KH95mod) values would instead produce L$_{bol}$ estimates lower than ours for objects with $M_\star < 0.4~M_\odot$, and higher than ours for objects with $M_\star > 0.8~M_\odot$. As a net result, the average age estimated for the cluster would not be significantly different from the one we report here; on the other hand, the shape of the data point distribution on the HR diagram would appear slightly modified with respect to that shown in Fig.\,\ref{fig:HR_models}, with the bulk of low-mass objects being shifted downward and the bulk of high-mass objects being shifted upward by $\lesssim$1~Myr along the model tracks.

Spatial coordinates, photometric data, effective temperatures and spectroscopic paramaters for the full sample of GES targets in the NGC~2264 field is reported in Table~\ref{tab:data}; the table also contains disk classification, accretion diagnostics and derived extinction and stellar parameters for the 655 cluster members identified among the GES sample and analyzed in this study.

\subsection{An age spread in NGC~2264?} \label{sec:age_spread}

The correspondence between the apparent luminosity spread observed on the HR diagram for young star clusters and an effective age spread among cluster members has long been debated in the literature. \citet{hartmann2001} listed a number of sources of observational error (including distance, spectral type, extinction, photometric variability, unresolved binarity, and disk accretion) that may impact the positioning of young stars on the HR diagram. As discussed in \citet{jeffries2012}, such uncertainties likely account for only a fraction of the apparent spread in L$_{bol}$ (up to an order of magnitude at any $T_{\rm eff}$); therefore, the observational inference that objects of a given SpT in a given SFR exhibit a broad range of luminosities appears to be robust. However, assessing to what extent this spread in L$_{bol}$ translates to a real age spread among cluster members is a more difficult issue. A way to probe this matter is to investigate how the luminosity/age spread observed on the HR diagram relates to independent stellar properties that provide information on the evolutionary status of the cluster population. 

Several studies, including mapping the RV distribution of young stars across the cluster (\citealp{furesz06}, \citealp{tobin2015}; see also \citealp{gonzalez2017}), have reported that NGC~2264 consists of multiple subpopulations, distinct in spatial and velocity properties, possibly related to a variety of star formation epochs within the region. While noting these kinematic indications of a possible age spread among cluster members, we defer any analysis of the kinematic structure of NGC~2264 from GES RV data to a following paper (Sacco et al., in preparation).

\begin{figure}
\resizebox{\hsize}{!}{\includegraphics{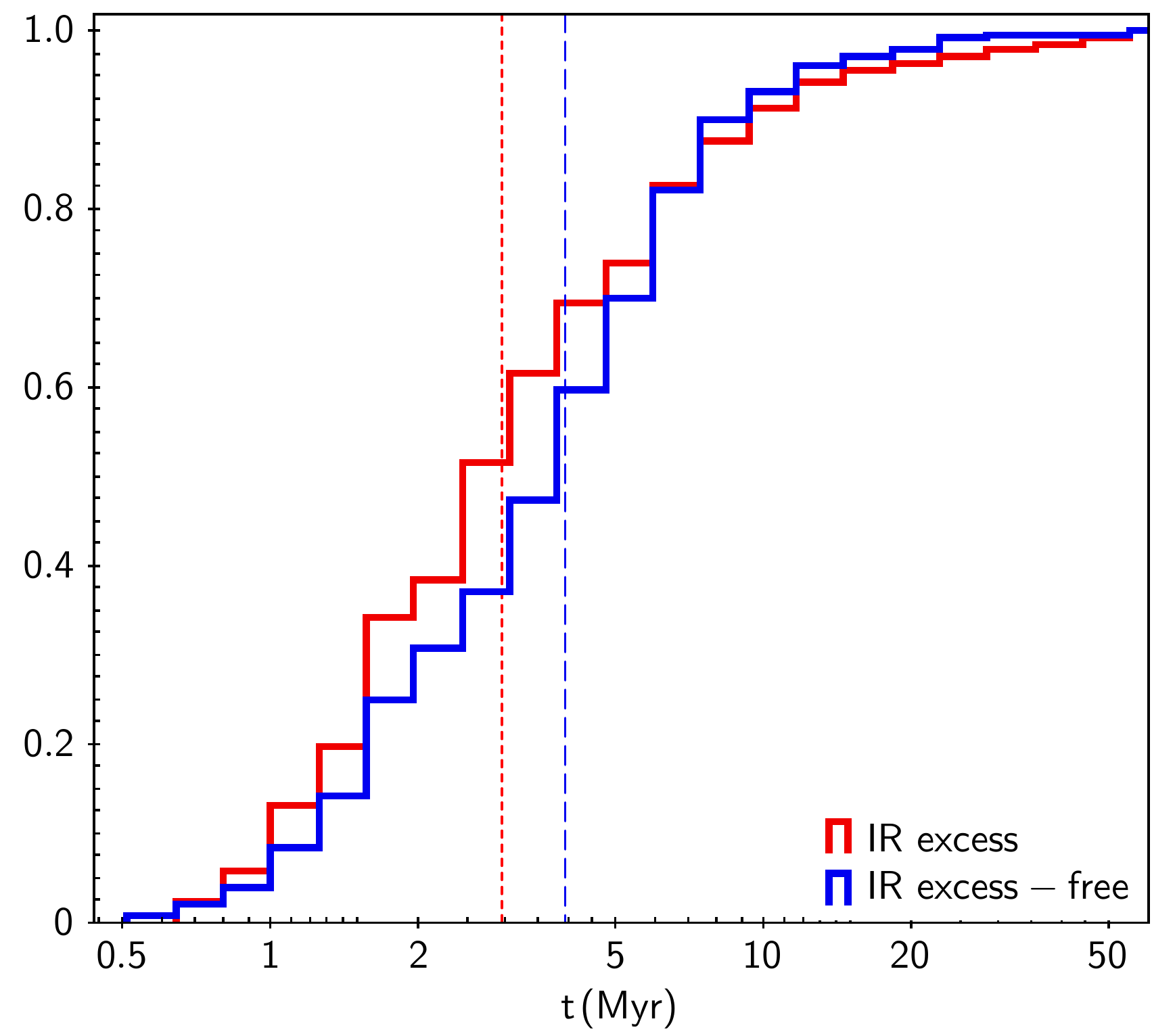}}
\caption{Normalized cumulative distributions in isochronal age for Class~II (red) and Class~III (blue) young stars in NGC~2264. The red dotted line and the blue dashed line mark, respectively, the median age measured across Class~II objects and that measured for Class~III objects in our sample.}
\label{fig:age_tts}
\end{figure}

In the assumption that all PMS objects follow a similar evolutionary pattern\footnote{This assumption is likely valid on a statistical level, although many different parameters, such as the disk mass, the disk viscosity, the X--UV luminosity from the central star, or the external radiation field, may play a role in determining the mechanisms and timescales of evolution of protoplanetary disks on an individual basis.}, from embedded (Class~I), to disk-bearing (Class~II), to disk-free (Class~III) \citep{lada1987}, we would expect the Class~II population of the region to be statistically less evolved (i.e., younger) than Class~III cluster members. Figure~\ref{fig:age_tts} illustrates the cumulative distributions in isochronal age associated with the disk-bearing (IR excess) and disk-free (no IR excess) populations of NGC~2264. 
Class~II and Class~III objects exhibit distinct median ages, of $2.93^{+0.37}_{-0.08}$~Myr and $3.93^{+0.32}_{-0.15}$~Myr \citep[cf.][]{gott2001}, respectively; this would support the suggestion that at least part of the age spread among cluster members deduced from the HR diagram is real. However, in both groups a significant dispersion of 0.3-0.4~dex is detected around the median age value, and the two distributions overlap substantially. A two-sample K-S test \citep{numerical_recipes} applied to the two distributions returns a $p$-value of 0.06 that they are extracted from the same parent distribution. This corresponds to the probability of obtaining a difference larger than that observed between the two cumulative distributions if these were representative of the same distribution. Hence, the null hypothesis that the distributions in age of Class~II and Class~III objects in our sample are statistically the same would be rejected at a significance level of 10\%, but would not be rejected at a significance level of 5\%. Similarly inconclusive is the same test applied to the populations of accreting and non-accreting sources in our sample: the median ages associated with the two groups would suggest that CTTS are on average younger than WTTS, in turn younger than Class~III stars, but the result is not significant at the 10\% level.

\citet{venuti2014} explored the spatial distribution of NGC~2264 members with ongoing accretion, and found a connection between their location within the cloud and the mass accretion rate ($\dot{M}_{acc}$) estimated from the UV excess detected for the sources. Namely, the strongest accretors appear to be projected onto the most embedded regions of NGC~2264, while more moderate accretors distribute more evenly across the cluster. Similar results were obtained by \citet{kalari2015} in the Lagoon Nebula ($\sim$1-2~Myr--old). Several studies \citep[e.g.,][]{hartmann1998, sicilia_aguilar2010} have reported that the average $\dot{M}_{acc}$ measured for CTTS in young clusters exhibits a definite decrease with stellar age $\left(\dot{M}_{acc}\,\, \propto\,\, t^{-1.5}\right)$; therefore, this association between the spatial location of accreting TTS within the region and their derived $\dot{M}_{acc}$ would indicate that the structure of the cluster reflects the formation history and dynamical evolution of its population. Objects of more recent formation appear to be predominantly located close to their birth sites, while more evolved objects distribute more evenly within a radius of a few parsecs around the innermost regions of NGC~2264.

\begin{figure}
\resizebox{\hsize}{!}{\includegraphics{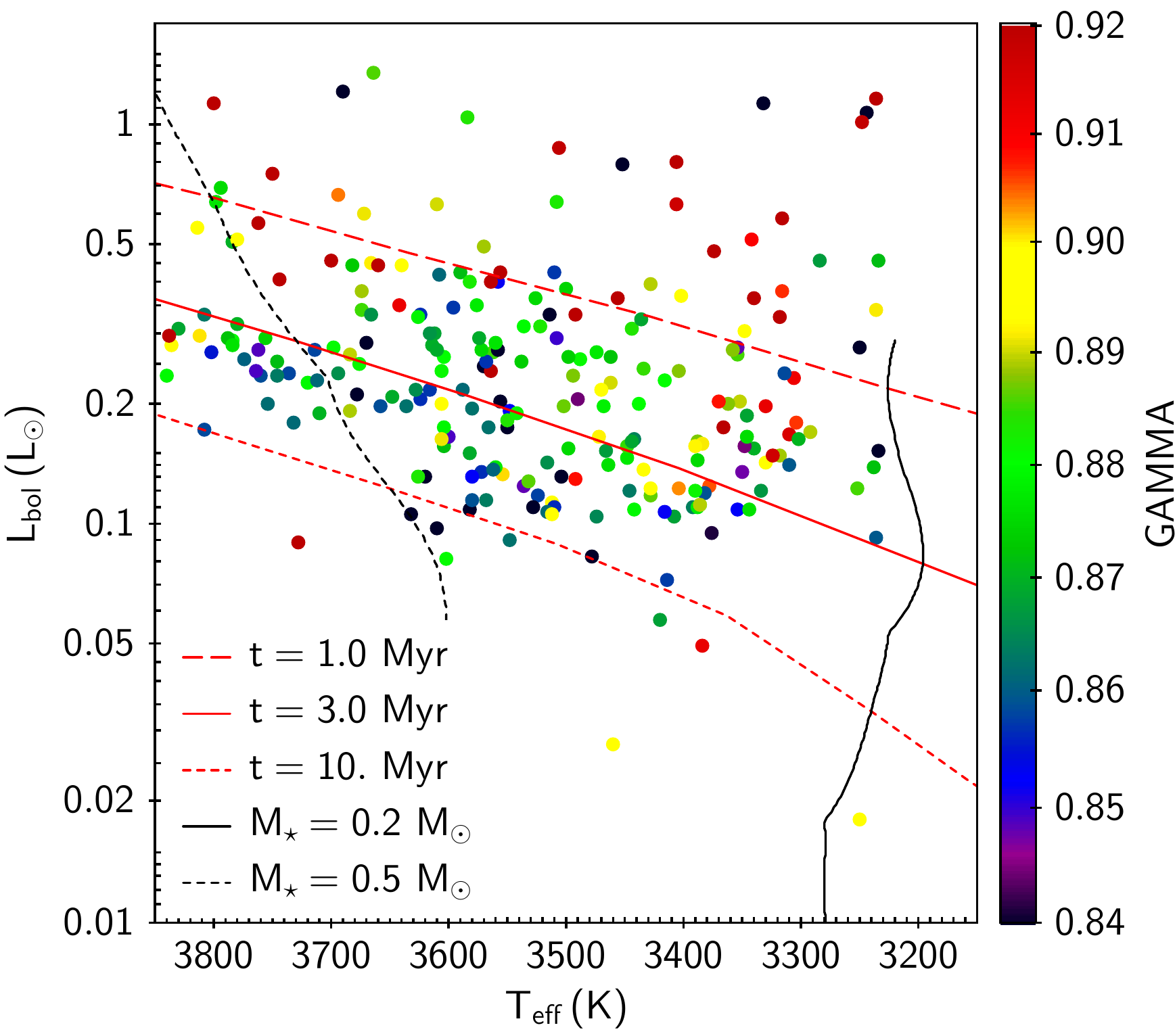}}
\caption{HR diagram of NGC~2264 members with $T_{\rm eff}$\,$<$\,3850~K (M-type stars). Colors are scaled according to their $\gamma$-index \citep{damiani2014}, as shown in the side axis. Mass tracks and isochrones traced on the diagram are from \citeauthor{baraffe2015}'s (\citeyear{baraffe2015}) PMS evolutionary models.}
\label{fig:HR_Mtype_gamma_zoom}
\end{figure}

Among the parameters released in GESiDR4, the $\gamma$-index \citep{damiani2014}, sensitive to stellar gravity, provides an empirical age indicator, completely independent of the photometric age measurements. Following its definition, higher values of $\gamma$ are indicative of lower gravities, which in turn reflect earlier evolutionary stages where the objects are still contracting. As illustrated in \citet{damiani2014}, the $\gamma$ enables rather clear discrimination between PMS and MS stars at a given $T_{\rm eff}$, and the average gravity properties derived for different young clusters in GES vary coherently with their relative age scale. In Fig.\,\ref{fig:HR_Mtype_gamma_zoom}, we test whether this parameter can also be applied to detect intra-cluster age differences that agree with the distribution of NGC~2264 members between the isochrones on the HR diagram. To this purpose, we have selected all objects in our sample with $T_{\rm eff}$ $<$ 3850~K, or equivalently spectral types later than M0 \citep{cohen1979, kenyon1995, herczeg2014}; this corresponds to the spectral range where the $\gamma$-index is independent of $T_{\rm eff}$ (see Figs.\,30 and 32 of \citealp{damiani2014}). 290 sources were retained following this selection, and are shown in Fig.\,\ref{fig:HR_Mtype_gamma_zoom} with colors that scale according to their $\gamma$.
A visual inspection of the diagram supports the idea that the vertical spread of cluster members at a given $T_{\rm eff}$ corresponds, at least statistically, to an actual age spread. Indeed, sources located in the upper part of the diagram, along lower-age isochrones, are typically associated with higher values of $\gamma$ (i.e., lower gravities); as we move from the top to the bottom of the data point distribution, or equivalently from the lowest-age to the highest-age isochrones on the diagram, the filling colors of the data points degrade from red (highest $\gamma$) to green and blue (lowest $\gamma$)\footnote{Similarly, \citet{dario2016} recently explored the correspondence between age estimates inferred from model-fitting on a HR diagram and those derived from model-fitting on a $\log{g}$ vs. $T_{\rm eff}$ diagram for young stars in Orion~A, and detected a significant correlation between the two sets of stellar ages.}. 

Strictly speaking, the gravity index is an indicator of stellar radius. Thus, the direct inference from Fig.\,\ref{fig:HR_Mtype_gamma_zoom} is that the distribution of objects at a given $T_{\rm eff}$ on the HR diagram statistically reflects a spread in stellar radii. A thorough discussion on the relation between spread on the HR diagram, dispersion in stellar radii and intrinsic age spread was presented in \citet{jeffries2007} for the slightly younger ONC cluster ($\sim$2.5~Myr). The author tested several scenarios, including a coeval population vs. a distribution in age, to interpret the nature of the derived dispersion in stellar radii at a given $T_{\rm eff}$ among ONC members, which exhibit apparent L$_{bol}$ and age spreads from the HR diagram comparable to those of NGC~2264 (Fig.\,\ref{fig:HR_models}). In the best-fitting scenario obtained for the ONC, external factors such as the observational uncertainties, while contributing a certain fraction to the apparent spread \citep{jeffries2011}, are not sufficient to explain its full extent: an additional, genuine age spread of a few Myr (i.e., comparable to, or larger than the median age of the cluster) is required to reproduce the dispersion of values. This picture may be consistent with what we observe here for NGC~2264.

To set our considerations on the $\gamma$-index on a more quantitative basis, we selected from Fig.\,\ref{fig:HR_Mtype_gamma_zoom} all objects located between the $t=1.0$~Myr and $t=3.0$~Myr isochrones (102), and those located between the $t=3.0$~Myr and $t=10.0$~Myr isochrones (93). We chose these two subgroups because they have similar statistical weight, and also to exclude the most extreme values of $\gamma$, prevailing above the $t=1.0$~Myr isochrone, that may be affected, for instance, by strong accretion activity (see discussion in \citealp{damiani2014}). The median $\gamma$ associated with objects in the $t=1-3$~Myr range is $0.880 \pm 0.002$, higher than the median $\gamma$ of $0.866 \pm 0.002$ computed across the $t=3-10$~Myr group. The difference of $0.014 \pm 0.003$ between the two values, corresponding to about half the interquartile range measured across the full sample in Fig.\,\ref{fig:HR_Mtype_gamma_zoom}, is non-negligible if we consider that \citet{damiani2014} found an average shift of only a few hundredths in $\gamma$ between PMS and MS stars in the same $T_{\rm eff}$ range in the field of the young cluster $\gamma$~Vel. A two-sample K-S test, applied to the two subpopulations extracted from Fig.\,\ref{fig:HR_Mtype_gamma_zoom}, returns a probability of only 0.004 that they are extracted from the same parent distribution in $\gamma$. This result indicates that NGC~2264 members distributed between the $t=3$~Myr and $t=10$~Myr isochrones are statistically more evolved (older) than those located at $t=1-3$~Myr. The result is even more clear if we consider the role of observational uncertainties on the individual stellar parameters shown in Fig.\,\ref{fig:HR_models}: in the assumption that individual uncertainties are unrelated from each other, these will tend to mix up the objects' true locations on the HR diagram, and therefore blur any existing trend among the data. Hence, our analysis suggests that, in spite of the issues on using model isochrones to derive individual ages for young cluster members, these can at least be adopted statistically to build a relative age scale to characterize the dynamics and evolution of the cluster population.

\section{Structure and star formation history of NGC~2264} \label{sec:structure_SFH}

The analysis reported in Sect.\,\ref{sec:age_spread} indicates that a real age spread is likely present among NGC~2264 members, as suggested by the HR diagram of the cluster and supported by evidence from independent properties of the objects such as their gravity or the presence/absence of disks and mass accretion. In the following, we use the isochrones from \citeauthor{baraffe2015}'s (\citeyear{baraffe2015}) models to statistically divide NGC~2264 members into different age bins, and investigate how Class~II, CTTS, Class~III, and WTTS members of different age distribute across the cluster.

\begin{figure}
\resizebox{\hsize}{!}{\includegraphics{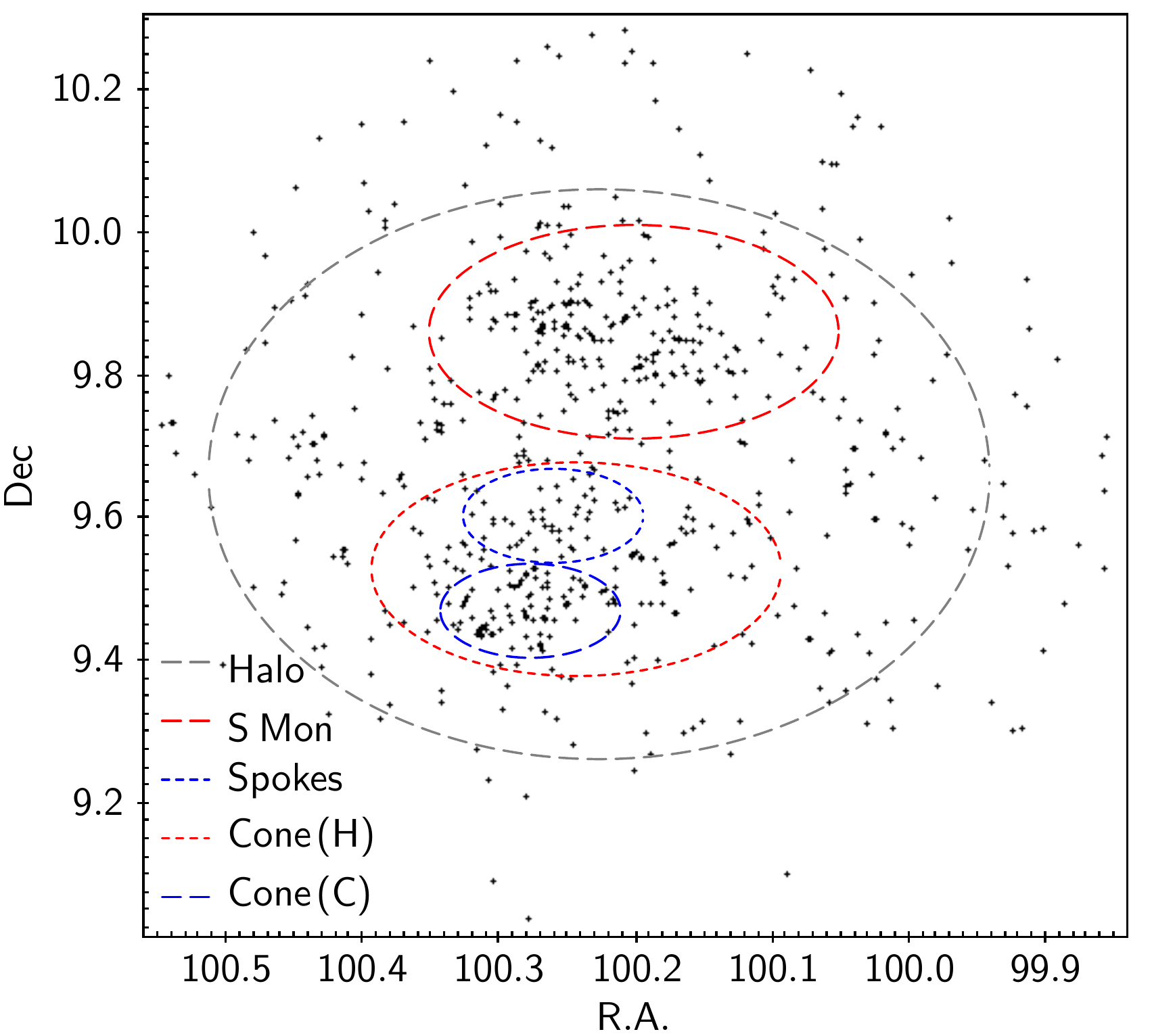}}
\caption{Spatial distribution of NGC~2264 members investigated in this study. The contours delimit the various subregions and subclusterings identified by \citet{sung09} across NGC~2264, as detailed in the legend in the bottom-left corner of the diagram. The area external to the Halo contour is referred to as Field.}
\label{fig:radec_Sung}
\end{figure}

\citet{sung09} analyzed {\it Spitzer} mid-IR photometry of NGC~2264 to map the spatial structure of the region and identify possible subclusterings. Based on the association between the spatial distribution of cluster members and their disk/envelope properties, the authors located two active SFRs within the cluster \citep[cf.][]{sung08}: S~Mon, in the northern half of NGC~2264, around the massive binary S~Mon, and Cone, in the southern half of the cluster, close to the tip of the Cone Nebula. These are surrounded by a Halo region where fewer cluster members are dispersed. In addition, \citet{sung09} distinguished two subregions inside the Cone area: the Spokes cluster, introduced in \citet{teixeira2006}, and Cone\,(C), the core of the Cone nebula region. These two embedded subclusterings contain most of the Class~I objects included in NGC~2264. The countours of the various NGC~2264 subregions listed in NGC~2264, overplotted to the R.A.--Dec diagram of cluster members investigated in this study, are shown in Fig.\,\ref{fig:radec_Sung}.

Here we adopt \citeauthor{sung09}'s (\citeyear{sung09}) scheme and explore what percentage of objects at different ages and evolutionary stages is located in each NGC~2264 subregion, to i) probe the formation and dynamical evolution of the cluster population, and ii) probe the impact of environmental conditions on star evolution across the cluster. To this purpose, we use isochronal ages as a statistical parameter to derive an age-dependent map of the cluster and investigate how its current structure has been shaped and evolved in time. A caveat for this analysis is that age estimates from isochrones are not independent of other stellar parameters; in particular, several studies \citep[e.g.,][]{hillenbrand2008, pecaut2016} have highlighted a spurious correlation trend between stellar mass and age which appears to be a common feature of different sets of PMS evolutionary models, although its extent varies from case to case. To appraise the potential impact of this effect on our analysis, we examined and compared the distribution in mass of objects at different spatial locations within the cluster. Details of this test are reported in Appendix~\ref{app:mass_dist}; in no case did we detect statistically significant distinctions in mass properties between the populations of different NGC~2264 subregions. We therefore surmise that the inferences of our study are not significantly affected by model-driven mass-dependent trends in isochronal ages.

\subsection{Nature of PMS stars across the cluster} \label{sec:PMS_radec}

\begin{table*}
\caption{Distribution of PMS objects in our sample across the six NGC~2264 subregions discussed in \citet{sung09}, and disk fraction, accretors fraction and distribution in age of the population of each subregion.}
\label{tab:tot_subreg_age}
\centering
\resizebox{\textwidth}{!}{%
\begin{tabular}{l c c c c c c c c c c c c c c c}
\hline \hline
\multirow{2}{*}{{\bf Region}} & \multirow{2}{*}{{\bf Tot \#}} & \multirow{2}{*}{{\bf Class~II (\%)}} & \multirow{2}{*}{{\bf CTTS (\%)}} & & \multicolumn{7}{c}{{\bf \# in each isochronal age bin (Myr)}} & & \multicolumn{3}{c}{{\bf \# with no age estimate in our sample}}\\
 & & & & & $<$\,1 & 1-2 & 2-3 & 3-4 & 4-6 & 6-8 & $>$8 & & $\swarrow$\,1\,Gyr\tablefootmark{\,a} & $\nwarrow$\,1.4\,M$_\odot$\tablefootmark{b} & \sout{($T_{\rm eff}$, L$_{bol}$)}\tablefootmark{\,c}\\
\hline
Spokes & 44 & 47.7$^{+18.2}$ & 52.3$^{+4.5}$ & & 7 & 8 & 4 & 3 & 3 & 1 & 4 & & 0 & 8 & 6\\
Cone\,(C) & 80 & 41.3$^{+12.5}$ & 37.5$^{+11.}$ & & 14 & 16 & 8 & 6 & 6 & 4 & 10 & & 0 & 3 & 13\\
Cone\,(H) & 70 & 28.6$^{+14.3}$ & 32.9$^{+4.3}$ & & 4 & 14 & 6 & 7 & 9 & 3 & 11 & & 1 & 4 & 11\\
S~Mon & 195 & 31.3$^{+11.8}$ & 23.6$^{+14.}$ & & 11 & 33 & 23 & 16 & 14 & 16 & 16 & & 2 & 10 & 54\\
Halo & 175 & 20.6$^{+10.9}$ & 22.3$^{+8.0}$ & & 6 & 27 & 14 & 21 & 29 & 22 & 16 & & 2 & 10 & 28\\
Field & 91 & 28.6$^{+12.1}$ & 28.6$^{+3.3}$ & & 0 & 5 & 6 & 8 & 9 & 14 & 8 & & 1 & 4 & 36\\
\hline
\end{tabular}
}
\tablefoot{ ``{\it Class~II~(\%)}'' = percentage of disk-bearing objects in the total number of sources located onto the corresponding region (the superscript corresponds to the percentage of objects with no IR information within the region, hence potentially adding to the estimated disk fraction). 
``{\it CTTS~(\%)}'' = percentage of accreting objects in the total number of sources located onto the corresponding region (the superscript corresponds to the percentage of objects with no UV/H$\alpha$ information within the region, hence potentially adding to the estimated accretors fraction).
\tablefoottext{a}{Objects whose location on the HR diagram falls below the 1~Gyr isochrone from \citeauthor{baraffe2015}'s (\citeyear{baraffe2015}) PMS model grid.}
\tablefoottext{b}{Objects whose location on the HR diagram falls above the highest mass track (1.4~M$_\odot$) included in \citeauthor{baraffe2015}'s (\citeyear{baraffe2015}) PMS model grid.}
\tablefoottext{c}{Objects that could not be positioned on the HR diagram because they lack an estimate of L$_{bol}$.}
}
\end{table*} 

In this Section, we combine isochronal ages with information on the evolutionary status of the objects (Class~II/Class~III, accreting/non-accreting) to characterize the nature of the cluster population from the most embedded to the most external regions. Table~\ref{tab:tot_subreg_age} summarizes how objects in our sample distribute among the various NGC~2264 subregions shown in Fig.\,\ref{fig:radec_Sung}, and how objects within each subregion distribute in age. Over 50\% of the objects are located onto the northern part (S~Mon) and the outer regions of the cluster. The disk fraction and the accretors fraction among cluster members are not uniform across the region: in the inner regions of the southern half of the cluster, the percentage of sources with active disks ranges from $\geq$40\% in Cone\,(C) to $\geq$50\% in Spokes; this percentage decreases to about 30\% in the outer shell of the Cone Nebula and in the northern half of the cluster (S~Mon), and to about 20\% in the Halo. The cumulative distributions of stellar ages associated with the most embedded regions of NGC~2264 (Spokes, Cone\,(C)) exhibit an excess of sources in the younger age bins compared to those associated with the more external regions of the cluster. To probe the nature and stellar content of each subregion further, we then examined the distribution of Class~II, Class~III, accreting (CTTS) and non-accreting (WTTS) objects in our sample (Sect.\,\ref{sec:disk_acc}) across the cluster, and computed the median isochronal age of each group of objects contained in each subregion separately. The results are presented in Table~\ref{tab:radec_all}.

\begin{table*}
\caption{Fraction and median isochronal age of Class~II, Class~III, accreting (CTTS) and non-accreting (WTTS) sources projected onto each of the six subregions of NGC~2264 discussed in \citet{sung09}.}
\label{tab:radec_all}
\centering
\begin{tabular}{l | c r l | c r l | c r l | c r l}
\hline \hline
 & \multicolumn{3}{c|}{\bf Class~II} & \multicolumn{3}{c|}{\bf Class~III} & \multicolumn{3}{c|}{\bf CTTS} & \multicolumn{3}{c}{\bf WTTS}\\
 & \multicolumn{1}{c}{Tot \#} & \multicolumn{1}{c}{\%} & \multicolumn{1}{c|}{Myr} & \multicolumn{1}{c}{Tot \#} & \multicolumn{1}{c}{\%} & \multicolumn{1}{c|}{Myr} & \multicolumn{1}{c}{Tot \#} & \multicolumn{1}{c}{\%} & \multicolumn{1}{c|}{Myr} & \multicolumn{1}{c}{Tot \#} & \multicolumn{1}{c}{\%} & \multicolumn{1}{c}{Myr} \\
\hline
Spokes & 21 & 10.7 & 1.71$_{-0.05}$ & 15 & 4.0 & 4.39$_{-1.08}$ & 23 & 12.3 & 1.77$_{-0.13}$ & 19 & 4.6 & 3.31$_{-0.52}$\\
Cone\,(C) & 33 & 16.8 & 1.91$_{-0.06}$ & 37 & 9.8 & 4.13$_{-0.49}$ & 30 & 16.0 & 1.54$_{-0.02}$ & 41 & 10.0 & 3.77$_{-0.76}$\\
Cone\,(H) & 20 & 10.2 & 3.96 & 40 & 10.6 & 3.45$_{-0.22}^{+0.08}$ & 23 & 12.3 & 4.43 & 44 & 10.8 & 2.86$_{-0.23}^{+0.04}$\\
S~Mon & 61 & 31.0 & 2.65$_{-0.20}$ & 111 & 29.4 & 3.72$^{+0.03}$ & 46 & 24.6 & 2.77$_{-0.05}^{+0.03}$ & 121 & 29.6 & 3.42$_{-0.06}^{+0.00}$\\
Halo & 36 & 18.3 & 3.85 & 120 & 31.8 & 4.16$_{-0.13}^{+0.08}$ & 39 & 20.9 & 3.90 & 122 & 29.8 & 4.00$_{-0.06}^{+0.14}$\\
Field & 26 & 13.2 & 6.43 & 54 & 14.3 & 4.60$^{+0.04}$ & 26 & 13.9 & 6.83 & 62 & 15.2 & 4.60$^{+0.04}$\\
\hline
\end{tabular}
\tablefoot{
``{\it Tot \#}'' = total number of sources in the corresponding category (Class\,II, Class\,III, CTTS, or WTTS) that are comprised within the corresponding NGC~2264 subregion.
``{\it \%}'' = percentage of sources in the corresponding category, with respect to the total number of NGC~2264 sources falling into that category, located onto the corresponding NGC~2264 subregion.
``{\it Myr}'' = median isochronal age associated with sources in the corresponding category in the corresponding NGC~2264 subregion.
Age estimates from our analysis are available only for sources that populate the area of the HR diagram delimited on the right by the 0.5~Myr isochrone, on the left by the 1~Gyr isochrone, and on the top by the 1.4~M$_\odot$ mass track (see Sect.\,\ref{sec:HR}). To take the first and second limit into account in our median ages, these are reported as $m^{+dm_2}_{-dm_1}$ (where applicable): $m$ is the median value of the measured isochronal ages in each subgroup; $m-dm_1$ is the computed median if we include objects that would appear younger than 0.5~Myr on the HR diagram; $m+dm_2$ is the computed median if we include objects that would appear older than the others on the HR diagram.
}
\end{table*}  

Independently of the group of objects being considered, a significant fraction of cluster members (about 50\% among Class~II/accreting sources, 60\% among Class~III/non-accreting sources) are located between the S~Mon and Halo regions, that is, in the northern half and in the more external areas of the cluster. The most embedded regions (Spokes and Cone\,(C) subclusterings) contain about 30\% of Class~II/accreting sources in our sample, and only about 15\% of Class~III/non-accreting sources. The median age associated with Class~II/accreting objects in the Spokes and Cone\,(C) subclusterings ($\sim$1.8 Myr) is younger than that of Class~II/accreting objects distributed across other regions of the cluster; this suggests that the former are of more recent formation, or that star formation activity has been ignited in the Spokes and Cone\,(C) subregions more recently than at other locations within the cloud. On the other hand, the median age of Class~III/non-accreting sources projected onto the Spokes and Cone\,(C) subregions is of the same order as that of Class~III/non-accreting sources distributed in other regions of the cluster; this may indicate that they are part of the Halo population of the region, and that their current position projected onto Spokes or Cone\,(C) is the result of post-birth migration from their formation sites located elsewhere in the cluster. The northern and the southern bulks of the cluster are distant a few parsecs from each other, consistent with the lengths that can be traveled by young stars over a few Myr assuming typical velocities of $\sim$1~km/s relative to the center of the stellar system (see, e.g., \citealp{bate2009}). \citet{sung2010} similarly reported an age gradient among the various subregions of NGC~2264, from the Spokes (the youngest) and Cone\,(C) subclusterings, to S~Mon and Cone\,(H), and finally to the Halo and Field regions (the oldest). To explain this age gradient, they suggested that star formation within NGC~2264 has occurred sequentially, starting from the outer envelope of the molecular cloud and propagating inwards, possibly ignited by an external source of triggering (e.g., supernova explosion). On the other hand, \citet{tobin2015} investigated the kinematic structure of NGC~2264 and identified a diffuse population of low-RV, blueshifted objects; further analysis by \citet{kounkel2016} showed that the blueshifted population is primarily located at southern declinations within the cluster, toward the Cone Nebula. \citet{tobin2015} suggested that these objects might be older than objects at higher RV (see also Sacco et al., in prep.), possibly the result of an episode of star formation that occurred inside the cloud even before the massive binary S~Mon was born; they may then have migrated outward to a halo encircling the inner regions of the cluster as a consequence of the photoionization and gas removal caused by S~Mon after its formation. A somewhat similar structure, layered in age, was inferred for the ONC by \citet{beccari2017}. In that case, the authors identified three separate populations within the cluster, all concentrated toward the cluster center but with different ages and rotational velocities. By examining their respective spatial distributions, \citeauthor{beccari2017} suggested that the youngest population of the ONC may be contained in the cluster core, surrounded by shells of older objects formed at earlier epochs.

In the outer regions of NGC~2264 (Halo, Field), the median age of cluster members is roughly uniform ($\sim$4--5~Myr) and similar across all stellar groups. Conversely, in the inner regions of the cluster (Spokes, Cone\,(C), S~Mon), the median age of Class~III objects is higher than that of non-accreting objects (WTTS), which is in turn higher than that of Class~II/CTTS. This suggests that the timescale relevant to the disk accretion activity is typically shorter than the disk lifetime, and shorter than the age of NGC~2264. Other studies conducted on young star clusters ({\small $\lesssim$}10~Myr; e.g., \citealp{bonito2013} for NGC~6611 and \citealp{sacco2008} for $\sigma$~Ori and $\lambda$~Ori), that reported populations of inert disks among cluster members, also support the view that the timescales relevant to disk accretion and to disk dissipation are distinct. \citet{fedele2010} adopted a multi-cluster approach to examine the time dependence of the fraction of accreting objects ($f_{acc}$) as opposed to that of disk-bearing objects ($f_{disk}$) in the age range from 2 to 30~Myr. They showed that $f_{acc}$ decreases more quickly with time than $f_{disk}$, and reported characteristic timescales of 2.3~Myr for disk accretion in young stars and of 3.0~Myr for dust dissipation in the inner disk; these estimates are consistent with the trends observed here for NGC~2264.

We note that a reverse trend in age between Class~II/CTTS and Class~III/WTTS objects would be deduced from Table~\ref{tab:radec_all} in the Cone\,(H) and, especially, Field regions. This result might be affected by selection effects. Table~\ref{tab:tot_subreg_age} illustrates that the Field population exhibits the highest rate of missing data in our age analysis among the various NGC~2264 subregions. We also note that, excluding the Spokes subcluster (which contains very few evolved sources), the Cone\,(H) and Field regions are those where the highest difference in median mass between Class~II and Class~III objects is registered (see Table~\ref{tab:KS_mass_tts}), although their respective mass distributions are not significantly different from each other according to a K-S test. In both regions, Class~II objects have larger median mass than Class~III objects, which might contribute to the fact that the former group appears comparatively older than the latter group (see Fig.\,\ref{fig:mass_age}).

\subsection{Constraining the dynamics and time evolution of the cluster}

In Sect.\,\ref{sec:PMS_radec}, we presented a snapshot of the current distribution of PMS objects across the main spatial extent of NGC~2264\footnote{We note, however, that recent studies have revealed a population of cluster members several parsecs away from the central star-forming sites; see, for example, \citet{venuti2014} and Salvaggio et al. (in preparation).}. We showed that the Halo surrounding the innermost cluster regions contains in proportion more Class~III (disk-free) than Class~II (disk-bearing) objects; the northern half of the cluster (S~Mon) encompasses similar percentages of Class~II and Class~III objects; the projected populations of the most embedded regions in the southern half of NGC~2264 (the Spokes and Cone\,(C) subclusterings) include a fraction of disk-bearing sources that is about twice as large as that of disk-free sources. The median ages associated with Class~II and Class~III objects in each subregion suggest that several episodes of star formation have occurred within NGC~2264. The subclusterings in the southern half of the region (close to the Cone nebula tip) appear to be the ones of most recent formation, $\sim$1.5--2~Myr ago; star formation activity may have started a few Myr earlier in the northern half of the cluster (around the massive binary S~Mon), and at the same time a population of PMS objects may have started to spread outward as a result of dynamical evolution from their birth sites. The numbers (median ages) reported in Table~\ref{tab:radec_all} suggest that a spread of 4-5~Myr exists among cluster members (see also \citealp{getman2014}). This is compatible with the age spread estimated by \citet{lim2016} from the Li abundances of NGC~2264 PMS objects (although \citealp{bouvier2016} examined the dispersion of EW(Li) for K4--M0 WTTS in NGC~2264 from GESiDR4 data and concluded that this scatter is likely not caused by an age spread).

\begin{figure*}
\centering
\includegraphics[width=\textwidth]{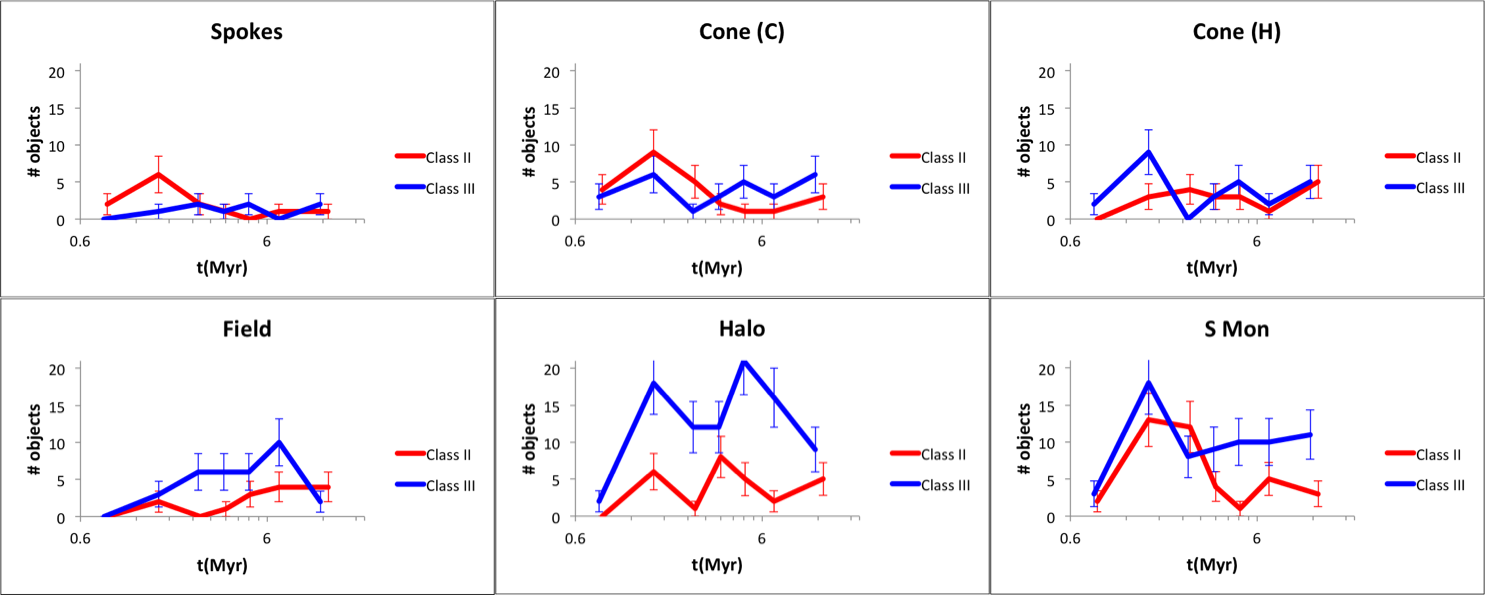}
\caption{Graphic representation of the data in Table~\ref{tab:radec_age}. Each panel illustrates the number of Class~II (red) and Class~III (blue) objects of a given age projected onto one of the six NGC~2264 subregions shown in Fig.\,\ref{fig:radec_Sung}. The x-axis coordinates correspond to the median isochronal ages computed for Class~II and Class~III objects in each age bin. The y-axis values can be derived from Table~\ref{tab:radec_age} by multiplying the total number of objects at a given age by the corresponding percentage in each region. The vertical error bars correspond to the Poissonian errors on the counts in each bin.}
\label{fig:sung_age_II_III}
\end{figure*}

In this Section, we attempt to retrace the evolution in time of the cluster population and structure. To this purpose, we restricted our analysis to objects classified as either Class~II or Class~III, as discussed in Sect.\,\ref{sec:disk_acc}, and that in addition have an individual age estimate derived following the procedure in Sect.\,\ref{sec:HR}. These amount to 391 stars, among which 138 Class~II and 253 Class~III objects. We then sorted these objects into 1 or 2~Myr-wide age bins, and we computed what percentage of Class~II and Class~III objects in each age group is projected onto each of the six NGC~2264 subregions shown in Fig.\,\ref{fig:radec_Sung}. The derived numbers are reported in Table~\ref{tab:radec_age}. Fig.\,\ref{fig:sung_age_II_III} illustrates the distribution in age of the Class~II/Class~III populations of each subregion; Fig.\,\ref{fig:radec_classII_III_evol} illustrates the spatial distribution of Class~II and Class~III objects in different age bins across the NGC~2264 field.

\begin{table*}
\caption{Distribution of Class~II and Class~III objects of different ages in the various subregions of the NGC~2264 cluster. }
\label{tab:radec_age}
\centering
\begin{tabular}{c | c c c c c c c | c c c c c c c }
\hline\hline
\multirow{3}{*}{\bf Myr} & \multicolumn{7}{c|}{\bf Class~II} & \multicolumn{7}{c}{\bf Class~III}\\
 & {\small tot} & {\small Spokes} & {\small Cone\,(C)} & {\small Cone\,(H)} & {\small S~Mon} & {\small Halo} & {\small Field} & {\small tot} & {\small Spokes} & {\small Cone\,(C)} & {\small Cone\,(H)} & {\small S~Mon} & {\small Halo} & {\small Field} \\
 &  & {\small (\%)} & {\small (\%)} & {\small (\%)} & {\small (\%)} & {\small (\%)} & {\small (\%)} &  & {\small (\%)} & {\small (\%)} & {\small (\%)} & {\small (\%)} & {\small (\%)} & {\small (\%)} \\
\hline
$<$1 & 8 & 25.0 & 50.0 & \,\,\,0.0 & 25.0 & \,\,\,0.0 & \,\,\,0.0 & 10 & \,\,\,0.0 & 30.0 & 20.0 & 30.0 & 20.0 & \,\,\,0.0\\
1--2 & 39 & 15.4 & 23.1 & \,\,\,7.7 & 33.3 & 15.4 & \,\,\,5.1 & 55 & \,\,\,1.8 & 10.9 & 16.4 & 32.7 & 32.7 & \,\,\,5.5\\
2--3 & 24 & \,\,\,8.3 & 20.8 & 16.7 & 50.0 & \,\,\,4.2 & \,\,\,0.0 & 29 & \,\,\,6.9 & \,\,\,3.4 & \,\,\,0.0 & 27.6 & 41.4 & 20.7\\
3--4 & 19 & \,\,\,5.3 & 10.5 & 15.8 & 21.1 & 42.1 & \,\,\,5.3 & 34 & \,\,\,2.9 & \,\,\,8.8 & \,\,\,8.8 & 26.5 & 35.3 & 17.6\\
4--6 & 13 & \,\,\,0.0 & \,\,\,7.7 & 23.1 & \,\,\,7.7 & 38.5 & 23.1 & 49 & \,\,\,4.1 & 10.2 & 10.2 & 20.4 & 42.9 & 12.2\\
6--8 & 14 & \,\,\,7.1 & \,\,\,7.1 & \,\,\,7.1 & 35.7 & 14.3 & 28.6 & 41 & \,\,\,0.0 & \,\,\,7.3 & \,\,\,4.9 & 24.4 & 39.0 & 24.4\\
$>$8 & 21 & \,\,\,4.8 & 14.3 & 23.8 & 14.3 & 23.8 & 19.0 & 35 & \,\,\,5.7 & 17.1 & 14.3 & 31.4 & 25.7 & \,\,\,5.7\\
\hline
\end{tabular}
\tablefoot{The total number of objects in a given age bin is reported in the ``tot'' column; the following columns indicate what percentage of Class~II/Class~III objects in the corresponding age bin is contained in each of the six subregions illustrated in Fig.\,\ref{fig:radec_Sung} (the values sum up to 100\% horizontally).
}
\end{table*}

From this analysis we can extract the following information.

\begin{enumerate}

\item At the youngest ages recorded ($\sim$0.8~Myr), objects with evidence of disks (Class~II) distribute among the most embedded regions of the cluster, where star formation is ongoing: the Cone\,(C), Spokes, and S~Mon subclusterings. The Cone\,(C) and S~Mon regions also encompass a significant fraction (60\%) of the Class~III objects in the same age bin; conversely, very few disk-free sources (8), at any age, are comprised in the population of the Spokes subclustering (Fig.\,\ref{fig:sung_age_II_III}, upper left panel). 

\item The two-dimensional selection of objects in the Spokes and Cone\,(C) subregions may be contaminated by sources located in the outer shell of NGC~2264, along the line of sight to the cluster. Indeed, a comparison of the average extinction properties of the various subregions (Table~\ref{tab:Av_comp}) shows that Spokes exhibits the highest median A$_V$ and, together with Cone\,(C), the largest A$_V$ dispersion; this suggests that their projected populations also comprise a component of low-extinction, non-embedded sources. An estimate of the expected number of contaminants to the true populations of Spokes and Cone\,(C) can be obtained by multiplying the number of objects in the Halo (Fig.\,\ref{fig:radec_Sung}) by the ratio between the area of Spokes/Cone\,(C) and the effective area of Halo. This amounts to about ten objects, that is, about 14\% of the sources projected onto Cone\,(C) and 25\% of the sources projected onto Spokes; these two subregions, therefore, might be even younger than they appear.

\item The most embedded regions of the cluster (Spokes, Cone\,(C)) become rapidly devoid of objects as the age window considered is shifted toward higher ages. The distributions in age of Class~II objects in Spokes (Fig.\,\ref{fig:sung_age_II_III}, upper left panel) and Cone\,(C) (Fig.\,\ref{fig:sung_age_II_III}, upper middle panel) exhibit a peak around t$\sim$1.5~Myr, followed by a clear decreasing trend. At t\,{\small $\gtrsim$}\,2.5~Myr, only a few percents (3.5\%) of objects are projected onto the Spokes subcluster, and almost no Class~II sources older than $\sim$2.5~Myr are found at these locations within the cluster. The absolute number of Class~III objects in Cone\,(C) appears to fluctuate around four at any age; however, the percentage of Class~III objects contained in Cone\,(C) with respect to the total population of Class~III sources at a given age drops from 30\% at t$\sim$0.8~Myr to about 10\% at t\,{\small $\gtrsim$}\,3.5~Myr (Table~\ref{tab:radec_age}). As discussed in the previous point, a contamination of a dozen sources is expected based on the population density of the Halo region; this may explain the apparent increase in number registered in the projected population of Cone\,(C) for ages t\,{\small $\gtrsim$}\,5~Myr.

\item At ages {\small $\gtrsim$}\,1--2~Myr, the bulk of the Class~II population shifts from the southern and innermost regions of NGC~2264 to the northern half (S~Mon) and the more external areas of the cluster. The Cone\,(H) (Fig.\,\ref{fig:sung_age_II_III}, upper right panel) and Field (Fig.\,\ref{fig:sung_age_II_III}, lower left panel) regions are relatively devoid of Class~II sources at any age; a peak occurs at t$\sim$2.5~Myr in Cone\,(H) and at t\,{\small $\gtrsim$}\,5~Myr in Field. The S~Mon region (Fig.\,\ref{fig:sung_age_II_III}, lower right panel) contains a significant fraction of Class~II sources in the age range $\sim$1--3~Myr; at later ages, however, it exhibits a sharp decline in number of disk-bearing members. This behavior is qualitatively different from what observed for the innermost regions (Spokes, Cone\,(C)), where a peak in population at very young ages is followed by a smoother tail of ``older'' sources; it is also different from what observed for the more external regions (Cone\,(H), Halo, Field), where the number of Class~II objects remains low at all ages and overall exhibits an increasing trend with age.

\item The highest fractions of Class~III sources are found in the S~Mon and Halo regions (lower right panel and lower middle panel of Fig.\,\ref{fig:sung_age_II_III}, respectively). The distribution in age of Class~III objects in S~Mon exhibits a rapid rise between t$\sim$0.8~Myr and t$\sim$1.5~Myr, and then settles around a uniform trend for t\,{\small $>$}\,2~Myr. A similar, albeit more fluctuating, behavior is observed for the Halo Class~III population. Each region encompasses between 20\% and 40\% of the total population of Class~III objects of the cluster across the whole age range. Except at the lowest ages sampled, only a small component of the Class~III population of NGC~2264 is found in the Cone region (Cone\,(H)+Cone\,(C)), merely a few percents at t$\sim$2.5~Myr.

\item Some trend inversions can be observed in the various panels of Fig.\,\ref{fig:sung_age_II_III} at the highest age bins (t\,{\small $\gtrsim$}\,7~Myr). This might be affected by the lower statistics available for higher ages as opposed to lower ages; moreover, in the case of Class~II objects, the presence of substantial material in the circumstellar environment might significantly dim the stellar luminosity, and therefore cause the object to appear older on the HR diagram than it actually is. 

\end{enumerate}

\begin{table}
\caption{Median A$_V$ and A$_V$ dispersion measured for cluster members projected onto each of the NGC~2264 subregions shown in Fig.\,\ref{fig:radec_Sung}.}
\label{tab:Av_comp}
\centering
\begin{tabular}{l c c}
\hline\hline
Subregion & Median A$_V$ & $\sigma$\tablefootmark{*} \\
\hline
Halo & 0.41 & 0.30 \\
S~Mon & 0.45 & 0.28 \\
Cone\,(H) & 0.44 & 0.36 \\
Cone\,(C) & 0.42 & 0.52 \\
Spokes & 0.58 & 0.82 \\
\hline
\end{tabular}
\tablefoot{\tablefoottext{*}{Resulting from an iterative 3\,$\sigma$-clipping routine.}
}
\end{table} 

The discussion reported above suggests the following.

\begin{itemize}

\item The Spokes subcluster is the youngest and most embedded region of NGC~2264; star formation activity at this location has presumably started only recently, a few Myr ago, less than the disk lifetimes, as indicated by the absence of Class~III objects.

\item The Cone\,(C) subcluster is also very young, as attested by the high percentage of Class~II sources of ages below 2~Myr contained in this region; however, the simultaneous presence of a significant population of Class~III objects indicates that star formation was ignited within Cone\,(C) earlier than in the Spokes subcluster.

\item The S~Mon region is likely older than Cone\,(C) and Spokes: the fraction of Class~II members it encompasses is highest for ages $\sim$2.5~Myr. This implies that star formation activity has started in the northern part of the cluster before it began in the southern part. At the same time, the presence of Class~II and Class~III objects that span the whole age range probed in our sample suggests that several episodes of star formation may have occurred, and that the subregion may still be actively forming stars.

\item No Class~II sources in the first age bin are located in the more external regions of the cluster (Cone\,(H), Halo, Field), which implies that no activity of star formation is currently ongoing in these regions. Interestingly, while the median ages associated with Class~III objects in the S~Mon and Cone\,(H) regions are very similar (Table~\ref{tab:radec_all}), the highest percentage of Class~II sources in Cone\,(H) occurs in the age bin around t$\sim$4.5~Myr, shifted by about 2~Myr toward older ages with respect to the location of the analogous peak observed for S~Mon.

\end{itemize}

\citet{sung2010} examined the median ages and cumulative age distributions of PMS stars in the various NGC~2264 subregions, to derive some information on the star formation processes within the region. The picture they deduced from their analysis is overall consistent with the scenario we outlined here: a picture of sequential star formation, where objects that populate the Halo and Field regions were formed first, followed by stars in S~Mon and Cone\,(H), then sources in Cone\,(C) and, finally, those in the Spokes subcluster. The separate analysis we conducted here for disked vs. non-disked (and accreting vs. non-accreting) objects as a function of age allowed us to: i) confirm the model-based indication of an age spread by the association with measurable quantities, such as the presence of disks or accretion; ii) obtain a more detailed overview of the dynamical evolution of the cluster; iii) explore the mechanisms and characteristic timescales of evolution of PMS objects across the region (see Sects.~\ref{sec:PMS_radec} and \ref{sec:photoevaporation}).

\subsection{Impact of environmental conditions on disk evolution} \label{sec:photoevaporation}

In their investigation of the star formation history of NGC~2264, \citet{sung2010} showed that, while the age distribution of the S~Mon population exhibits a significantly different behavior from those of the Spokes or Cone\,(C) populations (in the sense that the probability that they are drawn from the same parent distribution is very low), the populations of S~Mon and of Cone\,(H) exhibit very similar distributions in age. However, some differences between the two regions appear when we restrict our analysis to disk-bearing or accreting objects (see Table~\ref{tab:radec_all}). Namely, while the median ages associated with Class~III objects in S~Mon and in Cone\,(H) are similar to each other, the median age of disk-bearing objects in S~Mon is noticeably lower than that of disk-bearing objects projected onto Cone\,(H). The difference is more marked when only CTTS are considered: a K-S test applied to the age distributions of accreting objects in S~Mon and in Cone\,(H) returns a probability of 0.03 that they are drawn from the same population. 

This difference in median age between the two populations might be understood if objects in S~Mon were formed at a later time than objects in Cone\,(H). However, the fact that S~Mon hosts a significant fraction of Class~III objects at any given age, and the fact that the number of Class~II objects located in S~Mon decreases sharply after t$\sim$2.5~Myr (lower right panel of Fig.\,\ref{fig:sung_age_II_III}), would rather suggest that the characteristic timescales associated with disk accretion and disk evolution are shorter in the northern half of NGC~2264 than elsewhere within the region. This may result from non-uniform environmental conditions across the cluster.

The S~Mon region shown in Fig.\,\ref{fig:radec_Sung} is distinguished by the presence of the massive binary S~Mon (the primary component is an O7\,V star and the secondary component is an O9.5\,V star; \citealp{gies1993}). This is the only O-type star included in the population of the region, while another two dozen B-type stars are distributed within the cluster, and especially across the S~Mon and Cone regions \citep{dahm08}. Evidence of the impact that irradiation by nearby massive stars may have on the evolutionary pattern and timescales of young stars has been reported in the literature prior to this study. For example, \citet{guarcello2010} investigated the disk frequency across the Eagle Nebula, and found that this is lower in the most embedded regions of the nebula, at the location of the open cluster NGC~6611 ($\sim$2~Myr; \citealp{oliveira2008}, \citealp{bell2013}), where most OB stars of the region are concentrated, than in the outskirts of the nebula. A similar analysis was reported in \citet{guarcello2016} for the Cygnus OB2 association. This effect was tested in the case of NGC~2264 itself by \citet{sung09}: they calculated the disk fraction among cluster members as a function of the distance from the star S~Mon, and could detect a significant correlation between the two quantities (see their Fig.~17). On the other hand, \citet{richert2015} examined the issue of protoplanetary disk destruction by OB stars in six massive SFRs, and did not detect any evidence for inner disk depletion around young stars in the vicinity of O-type sources.

\begin{figure*}
\centering
\includegraphics[scale=0.63]{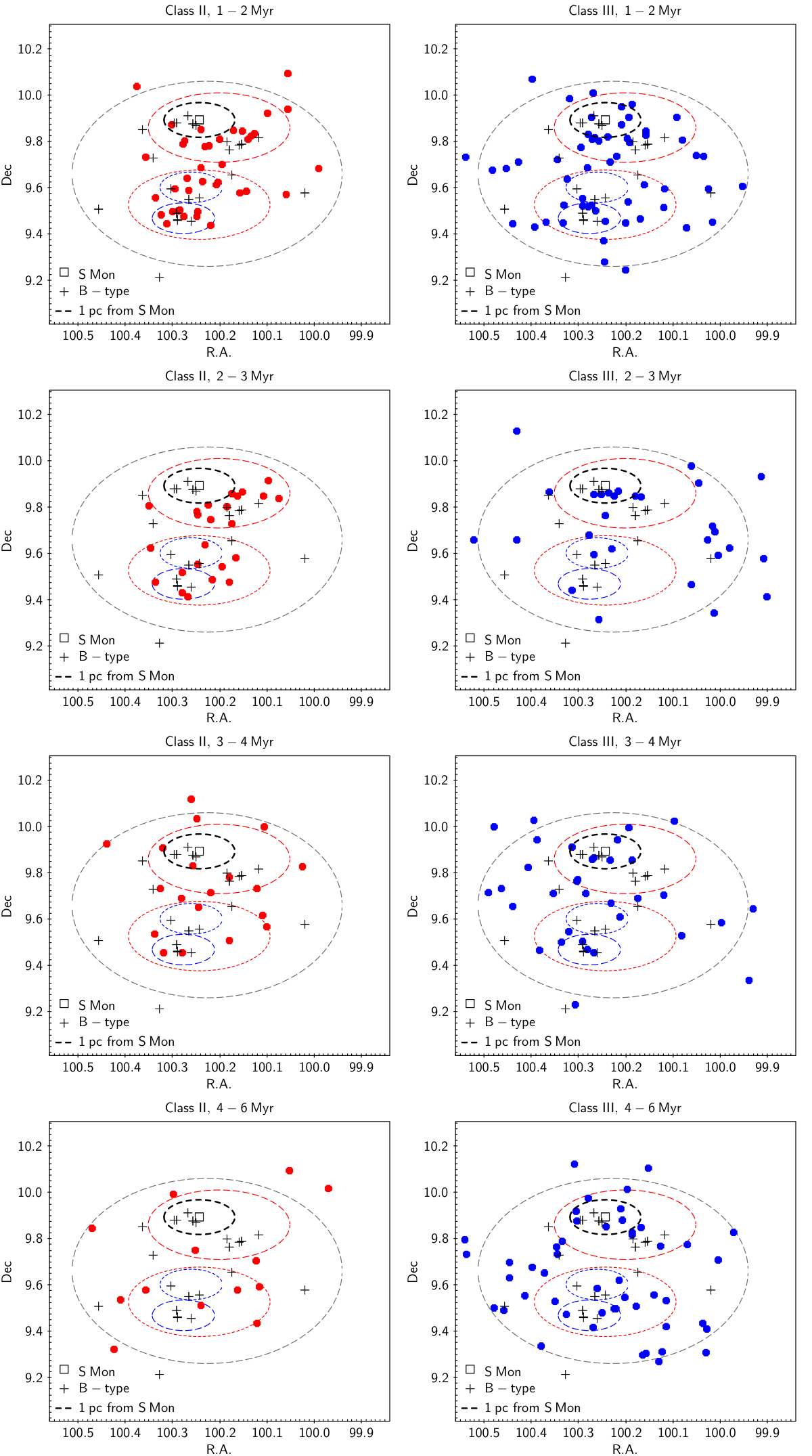}
\caption{Spatial distribution of Class~II (red dots, left panels) and Class~III (blue dots, right panels) NGC~2264 members in four age groups. The gray, red and blue contours highlight the cluster subregions shown in Fig.\,\ref{fig:radec_Sung}. The empty square identifies the O-type binary star S~Mon; the crosses mark the positions of other massive, B-type stars in the NGC~2264 region (from \citealp{dahm08}, Table~2). The black contour delimits the area within 1~pc from S~Mon.}
\label{fig:radec_classII_III_evol}
\end{figure*}

In Fig.\,\ref{fig:radec_classII_III_evol}, we illustrate the spatial distribution of Class~II and Class~III objects in selected age bins from Table~\ref{tab:radec_age} with respect to the positions of massive stars in the NGC~2264 field (listed in Table~2 of \citealp{dahm08}), and, in particular, relative to S~Mon. From the tables of \citet{martins2005} and \citet{parravano2003}, and assuming that the total luminosity of a binary star can be obtained as the sum of the luminosities of its components, we derive the following estimates for the EUV and FUV luminosity of S~Mon (O7V+O9.5V):
\begin{itemize}
\item $\log{L_{EUV}} = 48.7$~photons/s; 
\item $\log{L_{FUV}/L_\odot}=5.1$.
\end{itemize}
At a distance of 1~pc from the source, we thus have a FUV flux of 2522\,G$_0$ (where G$_0$ is the Habing flux, \citealp{habing1968}) and a EUV flux of $4.2 \times 10^{10}$ photons/s/cm$^2$. Although not extreme, such external FUV radiation fields can induce significant photoevaporation that, combined with the effect of internal viscosity, can effectively disperse a disk within 1--3~Myr \citep{clarke2007, anderson2013}. Following the prescription of \citet{adams2010}, the EUV flux estimate reported above would correspond to a photoevaporation-induced mass loss rate from the disk $\dot{M}_{EUV}=1.25\times 10^{-8} M_\odot$/yr, sufficient to disperse the disk in timescales of a few million years assuming a disk mass of order 0.01~M$_\odot$ \citep{garcia2011}. Therefore, this mechanism appears to be a plausible explanation for the evident dearth of older Class~II objects in the proximity of S~Mon, whereas Class~III sources populate the surroundings of massive stars within the cluster at any age (Fig.\,\ref{fig:radec_classII_III_evol}).

\section{Conclusions} \label{sec:conclusions}

We have conducted a study on the structure and dynamical evolution of the young cluster and SFR NGC~2264. This investigation is built primarily on the extensive spectroscopic (VLT/FLAMES) and photometric ({\it Spitzer}/IRAC, CFHT) observing campaigns that have targeted the region as part of the {\it Gaia}-ESO Survey and the Coordinated Synoptic Investigation of NGC~2264 projects. Our analysis encompassed 1892 objects in the NGC~2264 field, among which 655 were classified as cluster members based upon a variety of criteria including Li absorption, H$\alpha$ emission, IR excess, UV excess, X-ray emission. Cluster members in our sample are distributed in mass between 0.2 and 1.8~M$_\odot$; about 30\% of them exhibit a significant IR excess (Class~II), indicative of dusty disks, while another 58\% appear devoid of such an excess (Class~III) and, hence, disk-free. In addition, about 28\% of objects in our sample exhibit signatures of ongoing accretion (UV excess, H$\alpha$ emission), while data gathered for another 62\% qualify them as non-accreting (CTTS and WTTS, respectively).

The results of our study can be summarized as follows.

\begin{enumerate}

\item In agreement with previous studies, we find evidence of multiple substructures and subpopulations within the NGC~2264 regions. Less evolved sources (with significant amounts of material in the circumstellar environment) tend to be projected onto the most embedded regions of the cluster, in the north and close to the tip of the Cone Nebula, while more evolved sources distribute in a halo surrounding the innermost cluster regions.

\item We derive strong evidence for a real luminosity spread at a given $T_{\rm eff}$ on the HR diagram, which can be interpreted in terms of an age spread of $\sim$4~Myr among cluster members. This inference is supported by comparisons with other, independent stellar properties, such as the evolutionary state of the objects, and especially the gravity indicator $\gamma$-index, a new spectral index defined within GES and an effective empirical age indicator both at tracing a relative age ordering for star clusters and at detecting intra-cluster age spreads. 

\item The age spread and the structured spatial distribution of cluster members suggest that the PMS population of NGC~2264 is the result of multiple star formation events occurred sequentially over a few Myr. The widespread halo population of the cluster may have formed earlier in the innermost regions of the cluster and then migrated outward, possibly as a consequence of gas removal due to the ionizing action of NGC~2264 OB stars.

\item Star formation was not ignited at the same time throughout the region. It began in the northern part of the cluster, close to the location of S~Mon, and then propagated to the southern region of NGC~2264, around the Cone Nebula. The age estimates derived for cluster members suggest that the activity of star formation may have started over 5~Myr ago in the S~Mon subregion, and {\small $\lesssim$}\,1 Myr ago in the most embedded areas of the Cone Nebula subregion.

\item At the mean age of NGC~2264 ($\sim$3--5~Myr) it is already possible to distinguish between the characteristic timescales of disk accretion and those of disk dispersal. WTTS appear on average younger than Class~III objects across the inner regions of the cluster, which implies that the main disk accretion activity ceases before the inner disk is drained. The disk accretion phase is estimated to typically extend over 2--3 Myr.

\item The disk lifetimes among NGC~2264 members appear to be dependent on the specific location of the objects within the cluster. The Class~II to Class~III transition occurs at earlier ages in the northern half of the cluster (i.e., in the S~Mon subregion) than in the southern half of the cluster (i.e., in the Cone Nebula subregion). This likely reflects the impact of the energetic radiation from massive stars in the region (and most notably the O-type binary S~Mon) on rapid protoplanetary disk evolution and dispersal.  

\end{enumerate}

Our empirical knowledge of the nature and structure of young star clusters has profoundly increased over the last few decades thanks to extensive observational efforts aimed at mapping and characterizing the stellar content of clusters and SFRs in different environments, and to the concurrent modeling efforts to include these observations into realistic theories of star formation. A picture is emerging where young clusters often exhibit a hierarchical and fractal structure, with multiple subpopulations distinguished by age, spatial distribution, kinematic properties; this favors a sequential star formation scenario, or a multi-step molecular cloud collapse where several subclusters are formed and evolve somewhat independently before merging in the final, large-scale cluster. In the near future, the high-accuracy astrometric data that will be gathered for thousands of stars during the \emph{Gaia} mission, complemented with their kinematic characterization from the {\it Gaia}-ESO Survey, will provide us, for the first time, with complete three-dimensional maps of young clusters in our Galaxy, potentially marking a decisive contribution to our understanding of the star formation history in the Milky Way.

\begin{acknowledgements}
This work is based on data products from observations made with ESO Telescopes at the La Silla Paranal Observatory under program ID 188.B-3002. These data products have been processed by the Cambridge Astronomy Survey Unit (CASU) at the Institute of Astronomy, University of Cambridge, and by the FLAMES/UVES reduction team at INAF/Osservatorio Astrofisico di Arcetri. These data have been obtained from the {\it Gaia}-ESO Survey Data Archive, prepared and hosted by the Wide Field Astronomy Unit, Institute for Astronomy, University of Edinburgh, which is funded by the UK Science and Technology Facilities Council. This work is also based on observations made with the {\it Spitzer} Space Telescope, operated by the Jet Propulsion Laboratory, California Institute of Technology under a contract with NASA, and on data from MegaPrime/MegaCam, a joint project of CFHT and CEA/DAPNIA, at the Canada-France-Hawaii Telescope (CFHT) which is operated by the National Research Council (NRC) of Canada, the Institut National des Sciences de l’Univers of the Centre National de la Recherche Scientifique (CNRS) of France, and the University of Hawaii. The authors acknowledge useful discussions with S. Sciortino. We also thank V.~Kalari, A.~Klutsch, and L.~A.~Hillenbrand for their comments on an earlier version of this manuscript. This work was partly supported by the European Union FP7 program through ERC grant number 320360 and by the Leverhulme Trust through grant RPG-2012-541. We acknowledge the support from INAF and Ministero dell'Istruzione, dell'Universit\`a e della Ricerca (MIUR) in the form of the grant ``Premiale VLT 2012''. The results presented here benefit from discussions held during the {\it Gaia}-ESO workshops and conferences supported by the ESF (European Science Foundation) through the GREAT Research Network program. The authors acknowledge support through the PRIN INAF 2014 funding scheme of the National Institute for Astrophysics (INAF) of the Italian Ministry of Education, University and Research (``The GAIA-ESO Survey'', P.I.: S. Randich). S.H.P.A. acknowledges financial support from CNPq, CAPES and Fapemig. M.T.C. acknowledges financial support from the Spanish Ministerio de Econom\'ia y Competitividad, through grant AYA2016-75931-C2-1-P.
\end{acknowledgements}

\bibliographystyle{aa}
\bibliography{references}

\begin{appendix}

\section{Is the distribution in mass of cluster members uniform across the region?} \label{app:mass_dist}

\begin{figure}
\resizebox{\hsize}{!}{\includegraphics{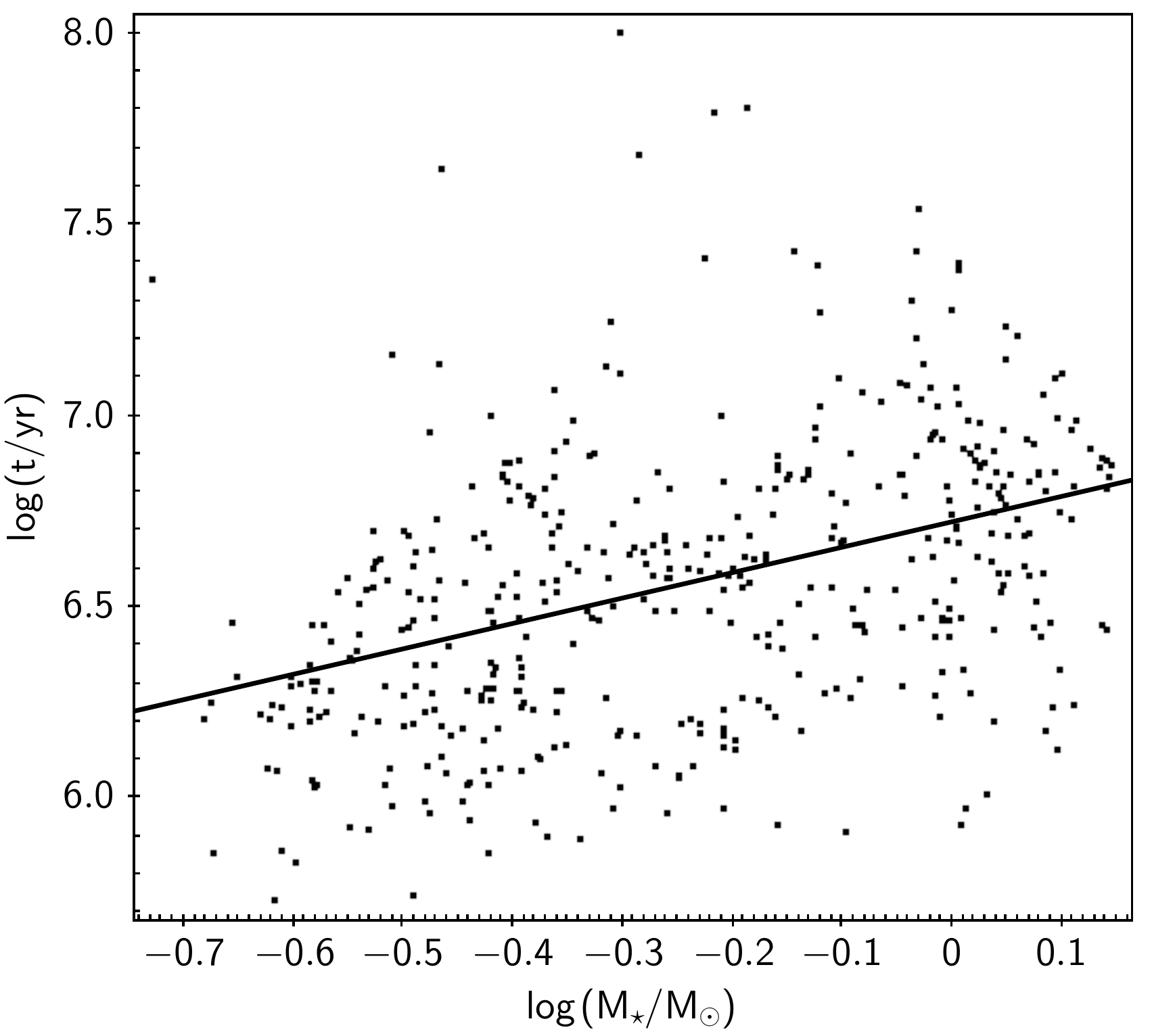}}
\caption{Comparison between stellar mass and age estimates derived for our NGC~2264 sample using \citeauthor{baraffe2015}'s (\citeyear{baraffe2015}) model grid. The solid line traces the linear correlation trend derived between the two variables ($r$ = 0.39; cf. \citealp{bevington}, Appendix~C.3).}
\label{fig:mass_age}
\end{figure}

Several studies \citep[e.g.,][]{hillenbrand2008, pecaut2016} have shown that lower-mass stars in a given young cluster tend to appear younger than higher-mass stars on a theoretical isochrone grid. This is likely an artificial feature due to an incorrect or incomplete treatment of convection and magnetic fields in lower-mass stars; different sets of models are affected by the same issue, as illustrated in Fig.\,3 of \citet{hillenbrand2008}. We tested whether this effect can be observed across our sample of parameters derived from the PMS evolutionary grid of \citet{baraffe2015}, and could indeed detect a significant ``correlation'' between stellar mass and isochronal age, as shown in Fig.\,\ref{fig:mass_age}.

\begin{figure}
\resizebox{\hsize}{!}{\includegraphics{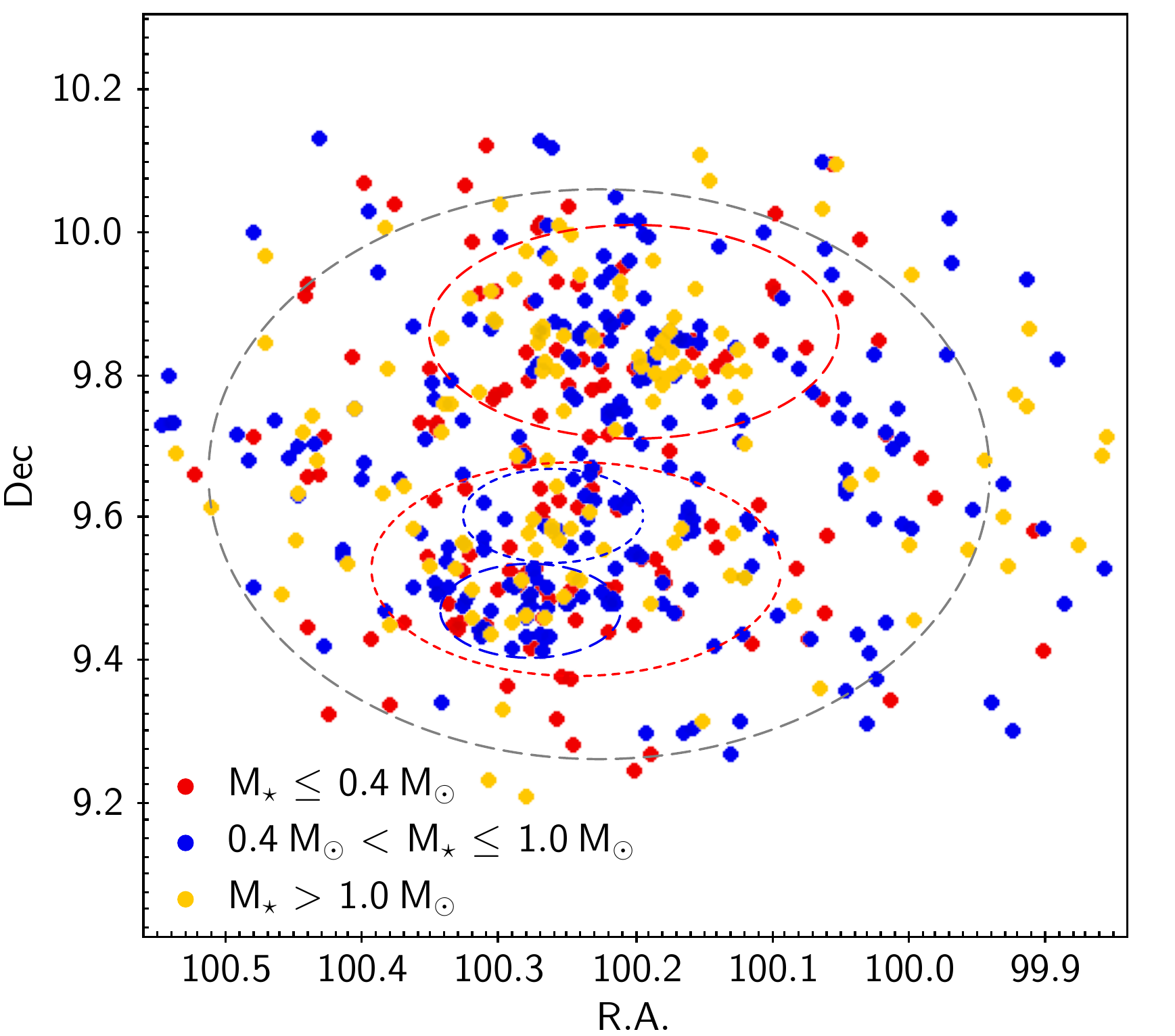}}
\caption{Spatial distribution of NGC~2264 members, sorted into three mass groups: $M_\star \leq 0.4\,\,M_\odot$ (red dots); $0.4\,\,M_\odot < M_\star \leq 1.0\,\,M_\odot$ (blue dots); $M_\star > 1.0\,\,M_\odot$ (yellow dots). The contours delimit the NGC~2264 subregions identified in \citet{sung09}: Spokes (blue dotted line), Cone\,(C) (blue dashed line), Cone\,(H) (red dotted line), S~Mon (red dashed line), Halo (gray dashed line), and Field (area external to the gray contour). Only objects with a mass estimate from the procedure discussed in Sect.\,\ref{sec:HR} are shown here.}
\label{fig:all_radec_mass}
\end{figure}

To assess the potential impact of this effect on our exploration of the spatial structure of the cluster, we conducted a comparative examination of the mass properties of objects at different locations across the region. In Fig.\,\ref{fig:all_radec_mass}, we illustrate the spatial distribution of our NGC~2264 member sample divided into three mass bins: $M_\star \leq 0.4\,\,M_\odot$; $0.4\,\,M_\odot < M_\star \leq 1.0\,\,M_\odot$; $M_\star > 1.0\,\,M_\odot$. This diagram shows qualitatively that there is no evident association between the spatial location of cluster members within the region and their mass properties: the three mass groups appear to be mixed across the whole region.  
To test this further on a quantitative basis, we divided our sample into six groups, each encompassing those objects that populate one of the six NGC~2264 subregions listed in \citet{sung09} and illustrated in Fig.\,\ref{fig:radec_Sung}: Field, Halo, Cone\,(H), S~Mon, Cone\,(C), and Spokes. We then considered these subgroups in pairs; for each pair, we run a two-sample, two-sided K-S test to compare the cumulative distributions in mass of objects in the two subregions and determine the probability that they are extracted from the same parent distribution (null hypothesis). The median masses computed in each region and the results of the K-S test are listed in Table~\ref{tab:KS_mass};
\begin{table*}
\caption{Median masses of the population of each NGC~2264 subregion listed in \citet{sung09}, and $p$-values resulting from applying a two-sample K-S test to the cumulative distributions in mass of each pair of subregions. }
\label{tab:KS_mass}
\centering
\begin{tabular}{l | l | c c c c c c}
\hline\hline
 & \multirow{2}{*}{\large \bf M$_\star$(M$_\odot$)} & \multicolumn{6}{c}{\large \bf $p$--values from two-sample K-S test}\\
 & & Field & Halo & Cone\,(H) & S~Mon & Cone\,(C) & Spokes \\
\hline
Field & 0.73 & & $\cdots$ & $\cdots$ & $\cdots$ & $\cdots$ & $\cdots$ \\
Halo & 0.54$_{-0.02}^{+0.01}$ & 0.112 & & $\cdots$ & $\cdots$ & $\cdots$ & $\cdots$ \\
Cone\,(H) & 0.63$_{-0.01}^{+0.03}$ & 0.728 & 0.586 & & $\cdots$ & $\cdots$ & $\cdots$ \\
S~Mon & 0.69$_{-0.04}^{+0.02}$ & 0.974 & {\it 0.083} & 0.629 & & $\cdots$ & $\cdots$ \\
Cone\,(C) & 0.55$_{-0.04}^{+0.05}$ & 0.283 & 0.932 & 0.675 & 0.162 & & $\cdots$ \\
Spokes & 0.70$_{-0.04}^{+0.08}$ & 0.607 & {\it 0.064} & 0.449 & 0.440 & 0.272 & \\
\hline 
\end{tabular}
\tablefoot{The upper and lower error bars associated with the median masses delimit the range of variability of the computed median if we take into account the objects with ($T_{\rm eff}$, L$_{bol}$) coordinates but no mass estimate from the HR diagram in Fig.\,\ref{fig:HR_models} because they fall outside the area covered by the model grid. When no error bars are reported, the variation on the value estimated following this approach is lower than 0.01. $p$-values smaller than 0.1 are reported in italic. The K-S test table is symmetric with respect to its diagonal.
}
\end{table*}
all $p$-values derived are larger than 0.05, which indicates that in no case the null hypothesis is rejected at a significance level of 5\%. This indicates that the distributions in mass of objects projected onto different NGC~2264 subregions are not significantly different from each other. There are two cases shown in Table~\ref{tab:KS_mass} (Halo vs. S~Mon and Halo vs. Spokes, in italic in the Table) where the $p$-value resulting from the two-sample K-S test stands between 0.05 and 0.1, which implies that the null hypothesis would be rejected at the 10\% significance level. A visual inspection of the cumulative distribution in mass of objects populating the Halo as opposed to those of the S~Mon and Spokes subclusterings indicates that the fraction of more massive stars contained in Halo is lower than that in S~Mon or Spokes (as also suggested by the respective median masses listed in Table~\ref{tab:KS_mass}). This may result in the S~Mon or Spokes populations seeming older than they actually are when we examine the isochronal ages relative to the Halo population (due to the effect shown in Fig.\,\ref{fig:mass_age}). 

Another test of relevance to our analysis consists in comparing the cumulative distributions in mass of disk-bearing and disk-free sources in each of the six NGC~2264 subregions shown in Fig.\,\ref{fig:radec_Sung}. If the null hypothesis is retained at each location, we can then perform a statistical comparison of the ages associated with objects at various evolutionary stages throughout the cluster, and probe its dynamical evolution. The results of this test are reported in Table~\ref{tab:KS_mass_tts}, that lists, for each subregion, the median masses of its Class~II and Class~III populations, and the probability of obtaining a difference larger than that observed between the cumulative distributions in mass of its Class~II and Class~III populations if these were randomly extracted from the same parent distribution.
\begin{table*}
\caption{$p$-values resulting from applying a two-sample K-S test to the distributions in mass of Class~II and Class~III objects projected onto each of the six NGC~2264 subregions shown in Fig.\,\ref{fig:radec_Sung}.}
\label{tab:KS_mass_tts}
\centering
\begin{tabular}{l l l l l l l}
\hline\hline
 & Field & Halo & Cone\,(H) & S~Mon & Cone\,(C) & Spokes \\
\hline
M$_\star$(M$_\odot$) [Class~II] & 0.76 & 0.55 & 0.70 & 0.65$_{-0.05}^{+0.13}$ & 0.59$_{-0.03}^{+0.02}$ & 0.66$_{-0.03}^{+0.02}$\\
M$_\star$(M$_\odot$) [Class~III] & 0.63$_{-0.01}^{+0.01}$ & 0.52$_{-0.02}^{+0.04}$ & 0.51$_{-0.02}^{+0.11}$ & 0.73$_{-0.01}$ & 0.50$_{-0.06}^{+0.15}$ & 0.96$_{-0.11}^{+0.12}$\\
$p$-value & 0.869 & 0.520 & 0.535 & 0.226 & 0.942 & 0.434 \\
\hline
\end{tabular}
\tablefoot{The upper and lower error bars associated with the median masses delimit the range of variability of the computed median if we take into account the objects with ($T_{\rm eff}$, L$_{bol}$) coordinates but no mass estimate from the HR diagram in Fig.\,\ref{fig:HR_models} because they fall outside the area covered by the model grid. When no error bars are reported, the variation on the value estimated following this approach is lower than 0.01. 
}
\end{table*}
These $p$-values are never smaller than 0.1, implying that the null hypothesis is never rejected to a significance of 10\%; we can therefore conclude that the distributions in mass pertaining to Class~II and Class~III objects at a given location within the cluster are not significantly different from each other.

We note that the tests discussed here were designed to assess whether biases deriving from a non-uniform distribution in mass of the cluster members included in our sample may have a significant impact on our analysis of the structure of the cluster as a function of isochronal age. A proper investigation on the issue of mass segregation within the cluster would need to encompass extensively all known cluster members, and in particular the more massive ones ($M_\star \geq 2-3\,\, M_\odot$), that are not included in the GES sample which is the focus of this study.

\section{Additional table}

\begin{sidewaystable*}
\caption{Complete list of the 1892 targets observed during GES in the NGC~2264 field, and of the measured spectroscopic parameters relevant to this study; CFHT optical photometry obtained for the same objects within CSI~2264; disk classification, accretion properties, and derived extinction and stellar parameters for the 655 cluster members identified among GES targets.}
\label{tab:data}
\centering
\resizebox{\hsize}{!}{%
\begin{tabular}{c c c c c c c c c c c c c c c c c c c c c}
\hline\hline
CNAME & CSIMon-\# & R.A. & Dec & {\small Memb} & $g$ & $r$ & $i$ & Class & Acc. & $\gamma$ & $T_{\rm eff}$ & {EW(Li)} & {EW(H$\alpha$)} & {W10\%(H$\alpha$)} & {$\Delta^{(r-H\alpha)}_{IPHAS}$} & {$E(u-r)$} & A$_V$ & L$_{bol}$ & M$_\star$ & Age\\
 &  & {(deg)} & {(deg)} &  &  &  &  &  &  &  & {(K)} & {(m\AA)} & {(\AA)} & {(km/s)} &  &  &  & {(L$_\odot$)} & {(M$_\odot$)} & {(Myr)}\\
\hline
06392396+0942016 & 006120 & 99.84983 & 9.70044 & & 17.370 & 14.993 & 13.745 & & & 1.046 & 4400 & {38} & & & & & & & & \\
06392408+0938088 & 012537 & 99.85033 & 9.63578 & & 19.672 & 17.960 & 16.982 & & & & & & & & & & & & & \\
06392497+0933151 & 006507 & 99.85404 & 9.55419 & & 18.417 & 17.020 & 15.805 & & & & & & & & & & & & & \\
06392506+0942515 & 006066 & 99.85442 & 9.71431 & {*} & 16.952 & 15.684 & 14.842 & III & N & 0.916 & 4704 & {492} & & {101} & {-0.08} & & 1.36 & 1.047 & 1.29 & 6.52\\
06392535+0943147 & 006028 & 99.85562 & 9.72075 & & 17.210 & 15.345 & 14.378 & & & 1.003 & 5126 & {34} & & & & & & & & \\
06392550+0931394 & 006491 & 99.85625 & 9.52761 & {*} & 18.226 & 16.746 & 15.808 & II & Y & 0.881 & 4272 & {497} & {5.8} & {411} & {0.03} & {-1.18} & 1.03 & 0.377 & 0.94 & 11.0\\
06392585+0938112 & 007916 & 99.85771 & 9.63644 & {*} & &  &  &  &  & 0.900 & 3607 & {548} & & & & &  &  &  & \\
06392623+0947066 & 012625 & 99.85929 & 9.78517 & & 18.213 & 17.105 & 16.478 & & & 1.009 & 5998 & {42} & & & & & & & & \\
06392625+0941108 & 001158 & 99.85937 & 9.68633 & {*} & 15.977 & 14.903 & 14.395 & III & N & 0.945 & 4907 & {513} & {1.6} & {225} & {-0.08} & {-0.11} & 1.15 & 1.354 & 1.39 & 6.90\\
06392649+0943298 & 005862 & 99.86037 & 9.72494 & & 18.448 & 16.897 & 16.016 & & & 1.032 & 5765 & {76} & & & & & & & & \\
\hline
\end{tabular}
}
\tablefoot{A full version of the Table is available in electronic form at the CDS. A portion is shown here for guidance regarding its form and content.\\
{\it ``CNAME''} = GES target identifier. 
{\it ``CSIMon-\#''} = Object identifier used within CSI~2264 \citep{cody2014}.
{\it ``R.A., Dec''} = Target coordinates.
{\it ``Memb''} = asterisks mark the cluster members selected among all GES targets as described in Sect.\,\ref{sec:memb_selection}.
{\it ``$g$, $r$, $i$''} = Optical photometry obtained with CFHT/MegaCam and calibrated to the SDSS system \citep{venuti2014}.
{\it ``Class''} = Object classification according to its disk properties (\citealp{lada1987}; II = disk-bearing; III = disk-free).
{\it ``Acc.''} = Object classification according to its accretion properties, following the criteria in Sect.\,\ref{sec:disk_acc} (Y = accreting, or CTTS; N = non-accreting, or WTTS).
{\it ``$\gamma$''} = GES gravity index \citep{damiani2014}; typical uncertainties are of order 0.006 for $\gamma \sim 0.85$ and 0.002 for $\gamma \sim 0.95$.
{\it ``$T_{\rm eff}$''} = GES released effective temperature; typical uncertainties are of order 3\% of the measurement.
{\it ``EW(Li)''} = GES released Li equivalent width; typical uncertainties range between 5 and 20 m\AA.
{\it ``EW(H$\alpha$)''} = GES released H$\alpha$ equivalent width; typical uncertainties range between 0.1 and 0.6 \AA.
{\it ``W10\%(H$\alpha$)''} = GES released H$\alpha$ width at 10\% intensity; typical uncertainties range between 10 and 20 km/s.
{\it ``$\Delta^{(r-H\alpha)}_{IPHAS}$''} = difference between the measured $r-H\alpha$ IPHAS color and the threshold traced in Fig.\,\ref{fig:IPHAS_accreting} to separate the accreting and non-accreting stellar loci; values significantly larger than 0 indicate H$\alpha$ emitters.
{\it ``$E(u-r)$''} = photometric UV excess measured in \citet{venuti2014} as the difference between the observed $u-r$ color and the reference $u-r$ color set by the WTTS population of similar brightness; the more negative the value of $E(u-r)$, the stronger the UV excess.
{\it ``A$_V$''} = Individual extinction estimate derived in this study.
{\it ``L$_{bol}$''}~= Bolometric luminosity derived in this study.
{\it ``M$_\star$''} = Stellar mass derived in this study.
{\it ``Age''} = Individual age estimated for the target using \citeauthor{baraffe2015}'s (\citeyear{baraffe2015}) model isochrones.
}
\end{sidewaystable*}

\end{appendix}

\end{document}